\documentclass[referee]{aa} % for a referee version
\usepackage{graphicx}
\usepackage{txfonts}
\begin{document}
\def\lsim{\,\lower2truept\hbox{${< \atop\hbox{\raise4truept\hbox{$\sim$}}}$}\,}
\def\gsim{\,\lower2truept\hbox{${> \atop\hbox{\raise4truept\hbox{$\sim$}}}$}\,}

   \title{Beam deconvolution in noisy CMB maps}

   \author{C.~Burigana\inst{1}
          \and
          D.~S\'aez\inst{2}
          }

   \offprints{C.~Burigana}

   \institute{IASF/CNR, Sezione di Bologna, 
              via P.~Gobetti, 101, I-40129 Bologna, Italy.\\
              \email{~burigana@bo.iasf.cnr.it}
             \and 
              Departamento de Astronom\'{\i}a y
              Astrof\'{\i}sica, Universidad de Valencia,
              46100 Burjassot, Valencia, Spain.~~~\email{~diego.saez@uv.es}
             }

   \date{Sent April 11, 2003}
%\date{Received September 15, 1996; accepted March 16, 1997}

   \abstract{The subject of this paper is beam deconvolution in small
angular scale CMB experiments. The beam effect is
reversed using the Jacobi iterative method, which was
designed to solved systems of algebraic linear equations.
The beam is a non circular one which moves according to the
observational strategy. A certain realistic
level of Gaussian instrumental noise is
assumed. The method applies to small scale CMB experiments in general
(cases A and B), but we have put particular attention on {\sc Planck} mission
at 100~GHz (cases C and D). In cases B and D, where noise 
is present, deconvolution allows to correct the main beam distortion effect
and recover the initial angular power spectrum 
up to the end of the fifth acoustic peak.
An encouraging result whose importance is analyzed in detail.
More work about deconvolution in the presence of other systematics
is in progress.

This paper is related to the {\sc Planck} LFI activities.

   }

\authorrunning{C.~Burigana \& D.~S\'aez}
\titlerunning{Beam deconvolution in noisy CMB maps}

   \maketitle
%
%________________________________________________________________

\section{Introduction}

Many experiments have been designed to measure Cosmic Microwave
Background (CMB) anisotropies at small angular scales.
Recent and new generation of experiments make use of 
multi-frequency and multi-beam instruments at a focus of a meter class
telescope. Since not all the feeds can be placed along the optical axis
of the telescope, the majority of them are necessary off-axis 
and the corresponding beams show more or less relevant optical
aberrations (see e.g. Page et al. 2003 for a recent discussion 
of the main beam shape in the context of the WMAP project),
according to the experiment optical design.
For example, in the contex of the ESA {\sc Planck} 
project~\footnote{http://astro.estec.esa.nl/Planck/} 
(see e.g. Bersanelli et al. 1996, Tauber 2000),
significant efforts have been carried out to significantly reduce
the main beam distortions produced by optical 
aberrations (see e.g. Villa et al. 1998, Mandolesi et al. 2000a).
On the other hand, even optimizing the optical design,
a certain level of beam asymmetry can not be 
completely eliminated (see e.g. Sandri et al. 2002, 2003). 

If not taken into account in the data analysis, the
main beam distortion introduces a systematic effect in the data
(Burigana et al. 1998, Mandolesi et al. 2000b) 
that affects the reconstructed map quality and, in particular, the 
recovery of the angular power spectrum of the CMB anisotropy (Burigana et al. 2001, 
Arnau et al. 2002, Fosalba et al. 2002).
The main topic of this paper is beam deconvolution
in this type of experiments with the aim of remove the main beam distortion
effect in the recovery of the angular power spectrum of 
CMB temperature anisotropy. We wish to reverse the
weighted average performed by a non-circular rotating beam
in the presence of a significant level of uncorrelated
instrumental noise.
The true angular power spectrum ($C_{\ell}$ quantities
before beam smoothing) should be recovered  from
the deconvolved maps, at least, in a large enough interval
($\ell_{min}$, $\ell_{max}$).

A preliminary work about beam deconvolution was presented in
Arnau \& S\'aez (2000).
In that paper, two methods for beam deconvolution were
considered.
One of them (hereafter method I) is based on the
Fourier transform, and the other one (method II) uses
the Jacobi algorithm for solving algebraic systems
of linear equations. Applications of these methods
in very simple cases were presented. The
first method was applied in the case of elliptical
non-rotating beams in the presence of a very low level
of Gaussian instrumental noise. The other method was only used
for a spherical beam in the total absence of noise.
More realistic situations must be considered. This
is the main goal of this paper.

The formalism of our approach to deconvolution is presented in Sect.~2.

Beam deconvolution can be only studied using simulations.
Map making requires a pixelisation, and
the accuracy of the angular power spectrum obtained from
pixelised maps strongly
depends on experiment sensitivity and resolution and on 
the sky coverage.
In the case of small angular scales ($\ell \gsim 100$),
the angular power spectrum can be well estimated using
small squared maps with edges lesser than $20^{\circ}$
(S\'aez et al. 1996).
In this case, a good map making algorithm and
an appropriate power spectrum estimator
were described in (Arnau et al. 2002).
In a first step (first part of Sect.~3), we work
with this type of simulations by assuming a simple elliptical 
main beam shape.
An observational 
strategy involving repeated measures in the same pixel but 
without a detailed reference to a specific experiment is adopted
at this stage. 

Afterwards, we apply the method to more complicated simulations 
carried out by using the 
HEALPix~\footnote{http://www.eso.org/science/healpix/}
({\it Hierarchical Equal Area and IsoLatitude Pixelization
of the Sphere}) package by G\`orski et al. (1999)
to pixelise the maps and compute the angular power spectrum 
from them. The beam size, its asymmetry, 
the variance of the instrumental noise,
and the beam rotation associated to an observational strategy
simulating the {\sc Planck} observations at 
100~GHz are considered in the second part of Sect.~3. 

Finally, we discuss our results and draw the main conclusions in Sect.~4.

We work in the framework of an inflationary flat universe 
(adiabatic fluctuations) with dark energy 
(cosmological constant) and dark matter (cold).
In this $\Lambda$CDM model, 
the
density parameters corresponding to baryons,  dark matter, and  
vacuum, are  $\Omega_{b} = 0.05$, $\Omega_{d} = 0.25$ and  
$\Omega_{\Lambda} = 0.7$,        
respectively, and the reduced Hubble constant is
$h=0.65$. No reionization is considered at all. All
the simulations are based on the CMB angular power spectrum 
corresponding to this model, which has been computed with 
the CMBFAST code 
by Seljak \& Zaldarriaga (1996).

\section{Beam smoothing and deconvolution}

Let us begin with an asymmetric non-rotating beam
which smoothes a map $T$ to give another map $T_{s}$.
In the continuous formalism, we can write:
\begin{equation}
T_{s}(\theta , \phi ) = \int B(\theta - \theta^{\prime},\phi -
\phi^{\prime}) T(\theta^{\prime} , \phi^{\prime}) d \Omega^{\prime}
\, ,
\label{conv}
\end{equation}
where the beam is described by function $B$.
If the beam centre points towards a point with
spherical coordinates ($\theta$, $\phi$), the
weight associated to another direction
($\theta^{\prime}$, $\phi^{\prime}$) is a function of
the form
$B(\theta - \theta^{\prime},\phi - \phi^{\prime})$.

In the absence of rotation, function $B$ only depends on the
differences $\theta - \theta^{\prime}$ and $\phi - \phi^{\prime}$ and,
consequently, Eq.~(\ref{conv}) is a mathematical convolution.
In such a case, the Fourier transform can be used (as it was explained
by Arnau and S\'aez 2000) to perform
beam deconvolution.
Nevertheless, if the asymmetrical beam rotates (as a result of
the observational strategy), the function describing the beam
is of the form
$ B = B(\theta - \theta^{\prime},\phi -
\phi^{\prime}, \theta, \phi )$.
Since the beam is
different (distinct orientations) when its centre points
towards different points in the sky,
a new dependence
on the angles  $\theta$ and  $\phi$ has appeared.
With a $B$ function involving this dependence,
Eq.~(\ref{conv}) is not a mathematical convolution and
the method I, which is based on the Fourier transform, does not work.

In practice, non-circular beams rotate due to the observational strategy
and, moreover, the effect of this rotation is not negligible in many
cases [see Arnau et al. 2002 for an estimation and Burigana et al. 2001 for 
an application to {\sc Planck} LFI (Low Frequency Instrument, 
Mandolesi et al. 1998)].
In this situation, method I cannot be used; however,
as we are going to show along the paper, method II works.

Let us assume a certain pixelisation and an asymmetric beam which
smoothes maps of the CMB sky. We first consider that only one 
observation per pixel is performed. The beam could have either the same
orientation for all the pixels or different orientations 
for distinct pixels; in both cases, 
the beam smoothes the sky temperature
$T$ to give $T_{s}$ according to the relation:
\begin{equation}
T_{s}^{i} = \sum_{i=1}^{M} B_{ij} T^{j}
\, ,
\label{disconv}
\end{equation}
where $M$ is the total number of pixels in the map. Quantity
$B_{ij}$ is the beam weight corresponding to pixel $j$
when the beam centre
points inside pixel $i$.
Eq.~(\ref{disconv}) can be seen as a system of $M$ linear
algebraic equations, in which, the independent terms $T_{s}^{i}$
are the observed temperatures, and the M
unknowns are the true sky temperatures $T^{j}$.
The Jacobi method can be tried in order to solve
this system. The solution would be the deconvolved map
with temperatures $T^{j}$.
No problems with pixel dependent beam orientations (rotation)
are expected {\em a priori}; nevertheless,
rotating asymmetric beams lead to a
$B_{ij}$ matrix which is more complicated than that corresponding
to a circular beam (studied in Arnau \& S\'aez 2000); by this reason,
we are going to verify
that Jacobi method works for any beam, first in the absence
of noise and, then, when there is an admissible noise level.
In matrix form, Eq.~(\ref{disconv}) can be written as follows:
$T_{s} = B T$.

If we now consider that each pixel is observed N times
either with an unique
beam and different orientations per pixel or with various non-circular
rotating beams (as it occurs in projects as {\sc Planck} where there are
various beams for each frequencies), then,
we can write N matrix equations (one for each measure)
of the form
$T^{(\alpha )}_{s} = B^{(\alpha )} T$, where index
$\alpha $ ranges from $1$ to $N$. The average temperature
assigned to pixel $i$ is $T^{i}_{a} = (1/N) \sum_{\alpha =1}^{N}
T^{(\alpha )i}_{s}$ and the above system of N matrix
equations leads to
\begin{equation}
T_{a} = B_{a} T \, ,
\label{nmea}
\end{equation}
where matrix $B_{a}$ describes the average beam, which
can be calculated as follows:
\begin{equation}
B_{a}^{ij} = \frac {1}{N} \sum_{\alpha =1}^{N} B^{(\alpha )ij} \, ;
\label{avbeam}
\end{equation}
hence, for a given experiment involving various measurements
per pixel, the average beam (\ref{avbeam}) might be calculated
at each pixel and, then, we can try to use the Jacobi
method to solve the system of linear equations~(\ref{nmea}).

We see that, in the absence of noise, method II could work in the
most general case, in which various beams move according to the most
appropriate observational strategy. 
For a multi-beam experiment one should in principle simulate
the effective scanning strategy and the convolution with the sky signal 
for each beam and then apply the formalism described above 
by taking into account the various resolutions and shapes 
of the beams. On the other hand, the power of this method is that 
it works independently of the small differences between the resolutions and
shapes of the various beams at the same frequency 
in a given experiment. Therefore, considering the 
data from a single average beam, instead of the data from 
the whole set of beams, 
but with the sensitivity per pixel obtained by considering
the whole set of receivers at the given frequency 
in the case in which the noise is taken into account, 
allows to reduce the amount of data storage and 
simplify the analysis without introducing a significant loss
of information about the accuracy of the method.

Instrumental uncorrelated Gaussian
noise makes beam deconvolution more problematic; nevertheless,
as we demonstrate in this work (Sects.~3.2 and 3.4), 
method II works for experiments with a 
level of noise similar to that of {\sc Planck}, or better, since 
the effect of deconvolution on the noise can be
quite accurately understood with Monte Carlo simulations.

\section{Applying method II}

In this section, the deconvolution method II is applied in
various cases using appropriate simulations. First, in cases A
and B, method II is applied to deconvolve a set of small
sky patches with regular pixelisation. 
To make this part of the work almost independent 
of the detailed scanning strategy 
of the considered CMB anisotropy experiment,
we adopted an observational strategy involving 
multiple observations of a given pixels and only roughly
mimicking that of {\sc Planck}.
The beam shape is assumed to be elliptical.
Case A does not involve any noise. 
Case B is identical to case A except for the presence of noise. 
Method II is then applied to deconvolve larger sky patches
but using the HEALPix package for the sky pixelization
and the computation of the angular power spectrum from
coadded and deconvolved maps, simulating 
the {\sc Planck} observational strategy, and assuming 
one of the beam shapes simulated in the past 
year for LFI, both in the absence of noise (case C) and in the presence
of noise (case D).

\subsection{Case A: noiseless, small patches}

An elliptical beam of the form
\begin{equation}
B(\theta-\theta^{\prime}, \phi - \phi^{\prime}) = W_{_{N}} e^ {\left[ -
\frac {(\theta - \theta^{\prime})^2}
{2 \sigma_{\theta}^{2}} - \frac {(\phi - \phi^{\prime})^2}
{2 \sigma_{\phi}^{2}} \right]}
\label{asb}
\end{equation}
is assumed, as in Burigana et al. (1998) and Arnau et al. (2002).

It is also assumed that quantities $\sigma_{\theta}$
and $\sigma_{\phi}$ obey the relations
$\sigma_{\theta} \sigma_{\phi} = \sigma^{2}$ and
$\sigma_{\theta}/\sigma_{\phi} = 1.3$, where $\sigma = 4.54^{\prime}$
($ \theta_{_{FWHM}} \simeq 10.68^{\prime} $). With this choice, the
elliptical beam (\ref{asb}) mimics the 100~GHz {\sc Planck} beams
for some locations of the detectors over the focal plane.

We simulate squared $14.6^{\circ} \times 14.6^{\circ}$ patches,
with 256 (128) nodes per edge; thus, our pixel size is
$\Delta = 3.43^{\prime}$ ($\Delta = 6.86^{\prime}$).
These sizes are allowed by
HEALPIX and, consequently, this choice will facilitate
some comparisons.
We use seventy five of these regions covering about the
forty per cent of the sky.
With this coverage and
$\Delta = 3.43^{\prime}$ ($\Delta = 6.86^{\prime}$),
the angular power spectrum can be estimated
(from simulated maps) with good accuracy from $\ell_{min} = 100$ to
$\ell_{max} = 10800/ \Delta \simeq 3100$ ($\ell_{max} \simeq 1550$).
See S\'aez \& Arnau (1997) for details about partial coverage.
In the theoretical model under consideration (see Sect.~1), 
the CMB temperature 
is a Gaussian homogeneous and isotropic statistical 
two dimensional field. In such a case,  
a certain method proposed by  
Bond \& Efstathiou (1987) can be used to make 
the $14.6^{\circ} \times 14.6^{\circ}$ maps 
used in this paper. This method is based on the
following formula: 
\begin{equation}
\frac {\delta T}{T} = \sum^{N}_{s_{1},s_{2} = -N} D(\ell_{1},
\ell_{2})e^{-i(\theta \ell_{1} + \phi \ell_{2})} \ ,
\label{mm}
\end{equation}
where $\ell_{1} = 2 \pi s_{1} / \Lambda$, 
$\ell_{2} = 2 \pi s_{2} / \Lambda$, and $\Lambda $ stands for  
the angular size of the square to be mapped.
This equation defines a Fourier transform from 
the position space ($\theta$, $\phi$) to the momentum space
($\ell_{1}$, $\ell_{2}$).    
The Gaussian quantities $D(\ell_{1},\ell_{2})$ have zero mean, 
and 
their variance is proportional to $C_{\ell}$, 
where $\ell = (\ell_{1}^{2} +
\ell_{2}^{2})^{1/2}$. Since $\delta T / T$ is real,   
the relation $D(- \ell_{1},- \ell_{2}) =D^{\ast}(\ell_{1},\ell_{2})$
must be satisfied. From given $C_{\ell}$ coefficients,  
the above  
$D(\ell_{1},\ell_{2})$ quantities  can be easily calculated and, then,
according to Eq. (\ref{mm}),
a Fourier transform leads to the map.
S\'aez et al. (1996) used this map making 
algorithm to get very good simulations of 
$20^{\circ} \times 20^{\circ}$ squared regions.   

In the case of small squared maps,
the above map making
method suggests the power spectrum estimator
used in Sects. 3.1 and 3.2 and also in Arnau et al. (2002). 
Given one of these maps $\delta T / T (\theta , \phi) $, 
an inverse Fourier transform leads to quantities 
$D(\ell_{1},\ell_{2})$ and, then, the average 
$\langle |D(\ell_{1},\ell_{2})|^{2} \rangle$
can be calculated on the circumference 
$\ell^{2} = \ell_{1}^{2} + \ell_{2}^{2}$. 
Some interpolations are necessary to get 
the $D(\ell_{1},\ell_{2})$ values at the points located on  
the circumference. 
The resulting average is proportional to $C_{\ell}$, where $\ell$
is the radius of the circumference.

%Simulations are made using the method proposed by
%Bond \& Efstathiou (1987) and tested by S\'aez et al. 1996.
%This method and the associated power
%spectrum estimator are described in detail by Arnau et al. (2002).  

For $\Delta = 3.43^{\prime}$ ($\Delta = 6.86^{\prime}$),
the beam average can be restricted to
a square with seventeen (nine) nodes per edge. Outside this
square, beam weights given by Eq.~(\ref{asb}) appear to
be negligible in this 
context~\footnote{The signal entering at angles 
larger than $\sim 1^\circ$ from the beam centre direction
produces the so-called straylight contamination, dominated
by the Galactic emission (see e.g. Burigana et al. 2001, 2003),
a systematic effect different from the main beam distortion
effect considered in this work.}.

Our elliptical beam is rotating while it covers
a given $14.6^{\circ} \times 14.6^{\circ}$ 
patch. In order to simulate beam rotation (see Fig.~1),
the squared patch is located with random orientation
(angle $\alpha $) in the plane ($\theta$ , $\phi $ ),
and a different beam 
orientation is assigned to each pixel of the patch. 
If Q is the centre of a certain pixel,
we find a point P on the $\theta $-axis which
is the centre of an auxiliary circumference with radius $85^{\circ}$
which passes by Q and, then, 
the beam orientation 
--~in the pixel under consideration~-- 
is fixed by assuming that
the major axis of the elliptical beam
is tangent --~at point $Q$~-- to the auxiliary circumference.
The distance from the patch centre, C, to the
$\theta $-axis is random, but it is constrained to be
smaller than $74^{\circ}$ in order to ensure the
existence of an auxiliary circunference passing by the centre of 
every pixel. 

   \begin{figure*}[t]
   \centering
   \includegraphics[width=18cm]{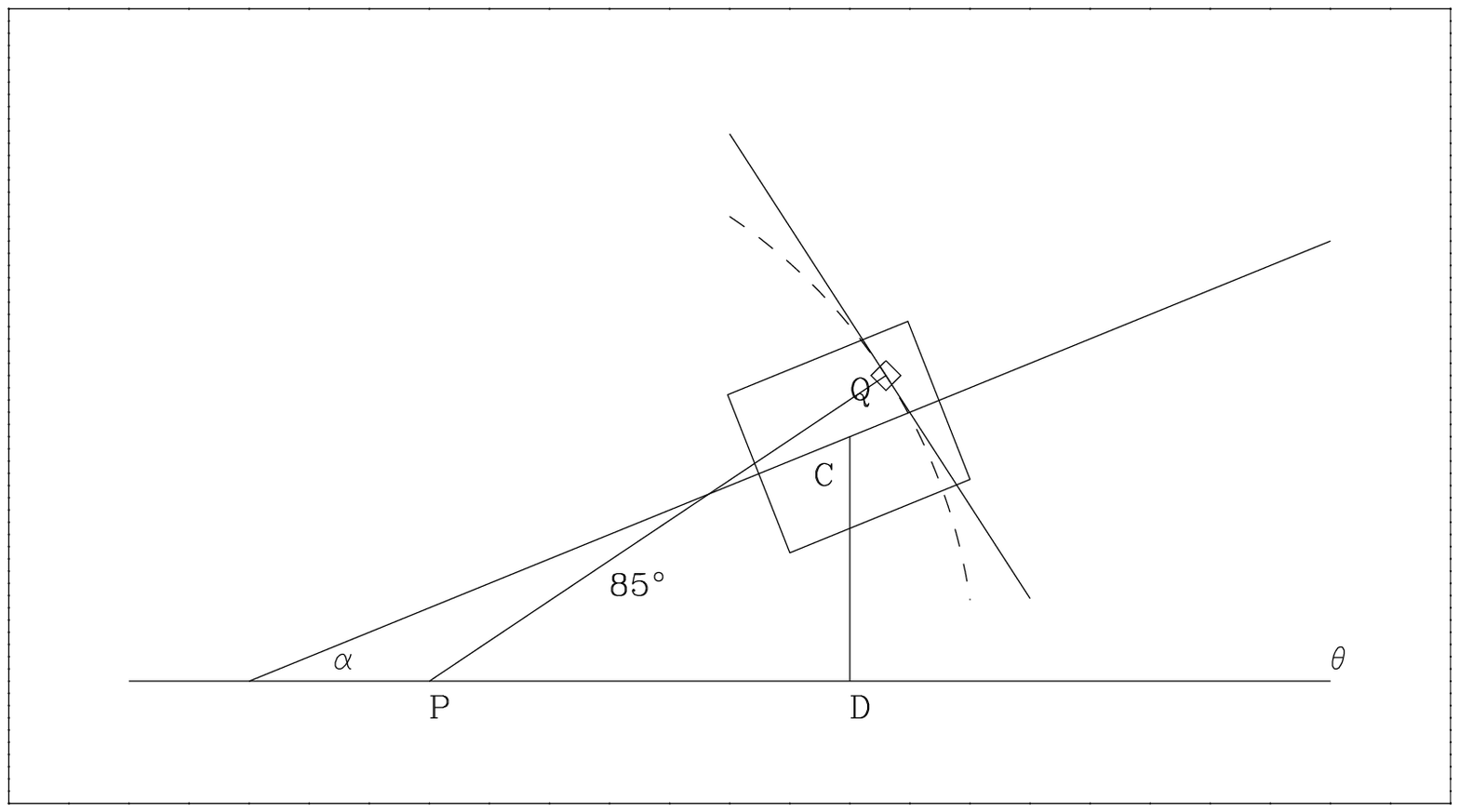}
   \caption{Patch location and beam orientation in cases
A and B defined in the text.
Points $C$ and $Q$ are the centres of
the patch and the pixel, respectively. Angle $\alpha $ and the
angular distance
$DC < 74^{\circ}$ define patch location.
Point $P$ is the centre of a circumference
with an angular radius of $85^{\circ}$, which passes by point Q.
The tangent to this curve
at $Q$ defines beam orientation (see also the text).}
    \end{figure*}

For each patch, the sky (T field) is simulated using
either 256 or 128 nodes per edge and, afterwards, the
beam described above is used to get the smoothed map ($T_{s}$).
The pixel temperatures of the $T_{s}$ map
are the independent terms of Eq.~(\ref{disconv})
and, moreover, the terms of the $B$ matrix can be built
up when necessary using Eq.~(\ref{asb}) and beam
orientation.
Taking into account that all the terms of the B diagonal
are identical to the central weight of the beam $b$, the $n+1$
iteration of the Jacobi method can be written as follows:
\begin{equation}
T^{(n+1)} = T^{(n)} + b^{-1}T_{s} - b^{-1}BT^{(n)} \, ,
\label{ja}
\end{equation}
where $T^{(n)}$ is the previous one. At zero order, we
take $T^{(0)} = T_{s}$.

Since the map is a $256 \times 256$ ($128 \times 128$) square
and the beam is another $17 \times 17$ ($9 \times 9$) square
(see Fig.~2), when the beam centre points towards 
a pixel located outside the ninth (fifth) row or column 
(counting from the nearest boundary), there are no
map temperatures to be weighted. In practice, for CMB maps,
we have verified that the following procedure works very well:
Write an equation for every internal node where the beam
average is well defined and, then,  solve the resulting system,
which has $ 240 \times 240$ 
($120 \times 120$) 
equations and the same unknowns.
The remaining temperatures (external points) are used when
required by beam smoothing,
but they are not altered along the iterative process.

   \begin{figure*}[t]
   \centering
   \includegraphics[width=7cm]{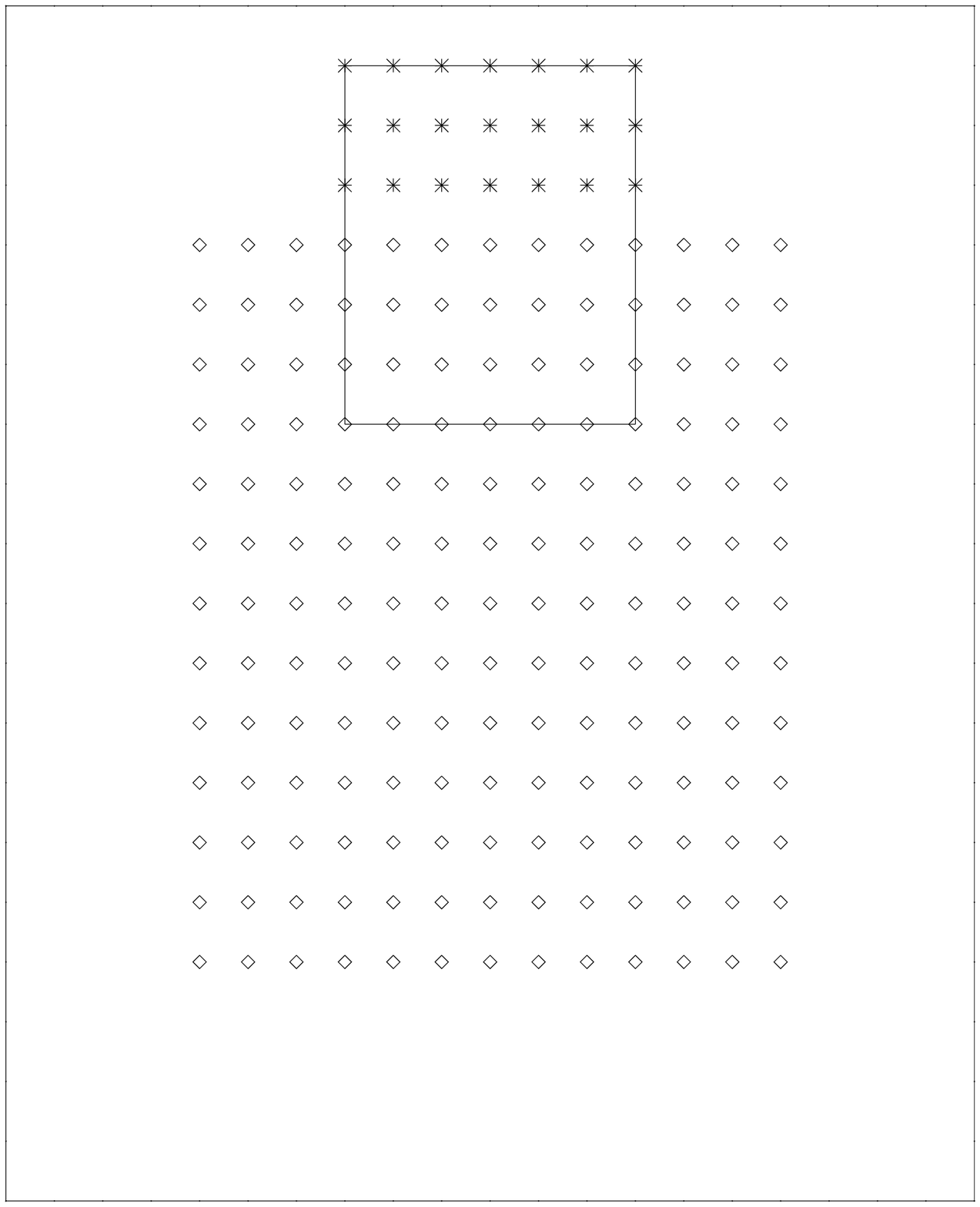}
   \caption{Boundary conditions for the application of the
Jacobi method in cases A and B (see text). An equation is written
at each internal node (diamonds). No equations are associated to
external points (asterisks). Temperatures of the external nodes keep
unaltered along the iterative process.}
    \end{figure*}

Fig.~3 shows the main results obtained in case A.
Top (bottom) left panel shows quantities
$\ell (\ell +1)C_{\ell} / 2 \pi$ in units of $\mu K^{2}$
before smoothing (continuous line) and after deconvolution
(pointed line)
for $\Delta = 3.43$ ($\Delta = 6.86$).
Both curves are indistinguishable
except at the largest $\ell $ values included
in the Figure. The relative deviations between
the dotted and dashed lines of each panel are given in
the corresponding right panels. The relative error introduced
by deconvolution --~in the absence of noise~-- is smaller than
$5 \%$ (0.5 \%) for $\ell \leq 1900$ ($\ell \leq 1480$)
in the case $\Delta = 3.43$ ($\Delta = 6.86$); 
the deviations grow beyond the sixth (fifth) acoustic peak.

   \begin{figure*}[t]
   \centering
   \includegraphics[width=18cm]{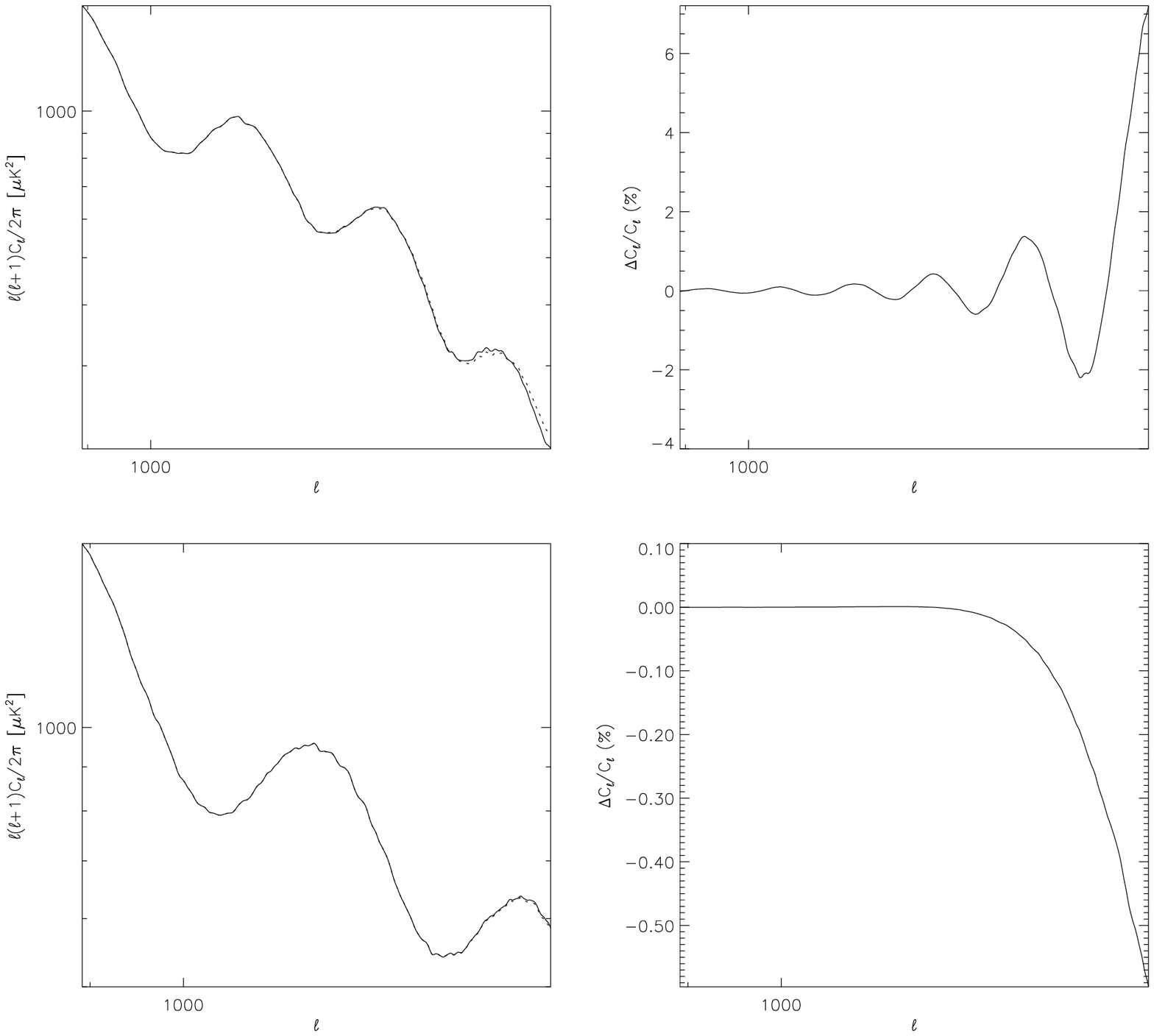}
   \caption{Top left panel displays quantities
$\ell (\ell +1)C_{\ell} / 2 \pi$, in units of $\mu K^{2}$,   
extracted from simulated maps
with a pixel size $\Delta = 3.43$.
solid (dotted) line has been obtained from unconvolved (deconvolved)
maps. Top right panel shows the relative deviations between the
spectra of the top left panel. Bottom
panels are the same as top ones but for $\Delta = 6.86$ (see also the text).}
    \end{figure*}

\subsection{Case B: noisy, small patches}

In this case, beam, rotation, coverage and pixelizations are
identical to those of case A; however, there is instrumental
uncorrelated Gaussian noise with $\sigma_{_{N}} = 9 \ \mu K$
($\sigma_{_{N}} = 4.5 \mu K$), in antenna temperature, 
for $\Delta = 3.43$ ($\Delta = 6.86$),
just the noise expected by combining all the beams of 
{\sc Planck} at 100~GHz.
A joint treatment of the impact of main beam distortions and of 
correlated $1/f^{\alpha}$ type noise 
(see e.g. Seiffert et al. 2002) and other kind of instrumental
systematics (see e.g. Mennella et al. 2002) is out of the scope
of this paper. On the other hand, this does not represents a 
crucial limitation, since blind destriping algorithms can strongly
reduce the impact of these effects (see e.g. Delabrouille 1998, Maino et al. 1999,
Mennella et al. 2002) also in the presence of optical distortions
(Burigana et al. 2001) and, possibly, of non negligible foreground
fluctuations (Maino et al. 2002).
The system to be solved has the form:
\begin{equation}
T_{s}^{i} = \sum_{i=1}^{M} B_{ij} T^{j}+N^{i}
\, ,
\label{noise}
\end{equation}
where $N^{i}$ is the noise at pixel $i$.
Using matrices, this equation can be written as follows:
\begin{equation}
T_{s} = B (T + B^{-1} N)
\, .
\label{triv}
\end{equation}
This last equation is formally identical to the matrix form
of Eq.~(\ref{disconv}) and it can be solved in the same way
--~using Jacobi method~-- to find the map
$T + B^{-1} N$. After applying this method, some
numerical error $E$ is expected and, consequently,
the numerical solution of system
(\ref{triv}) is of the form
\begin{equation}
T^{*} = T + B^{-1}N + E
\, .
\label{fal}
\end{equation}
In general, $T^{*}$ is different from $T$ (sky
temperature before smoothing); hence,
the angular power spectrum
extracted from the map $T^{*}$ is different from that
of the unconvolved sky, which is to be extracted from
map $T$.
Results are shown in Fig.~4, which has the same structure as
Fig.~3. We see that the spectra before smoothing and
after deconvolution
(which are obtained from maps $T$ and $T^{*}$, respectively)
separate at middle $\ell $ values. In the right panels, we can
verify that the deviation produced by deconvolution
--~in the presence of the assumed level of noise~-- is
smaller than
five per cent for $\ell \leq 1100$ ($\ell \leq 950$)
in the case $\Delta = 3.43$ ($\Delta = 6.86$). 
The relative deviations rapidly increases
for greater $\ell $ values. From the comparison of Figs.~3
and 4 it follows that the presence of noise with
$\sigma_{_{N}} = 9 \mu K$  ($\sigma_{_{N}} = 4.5 \mu K$)
has important consequences for beam deconvolution.

   \begin{figure*}[t]
   \centering
   \includegraphics[width=18cm]{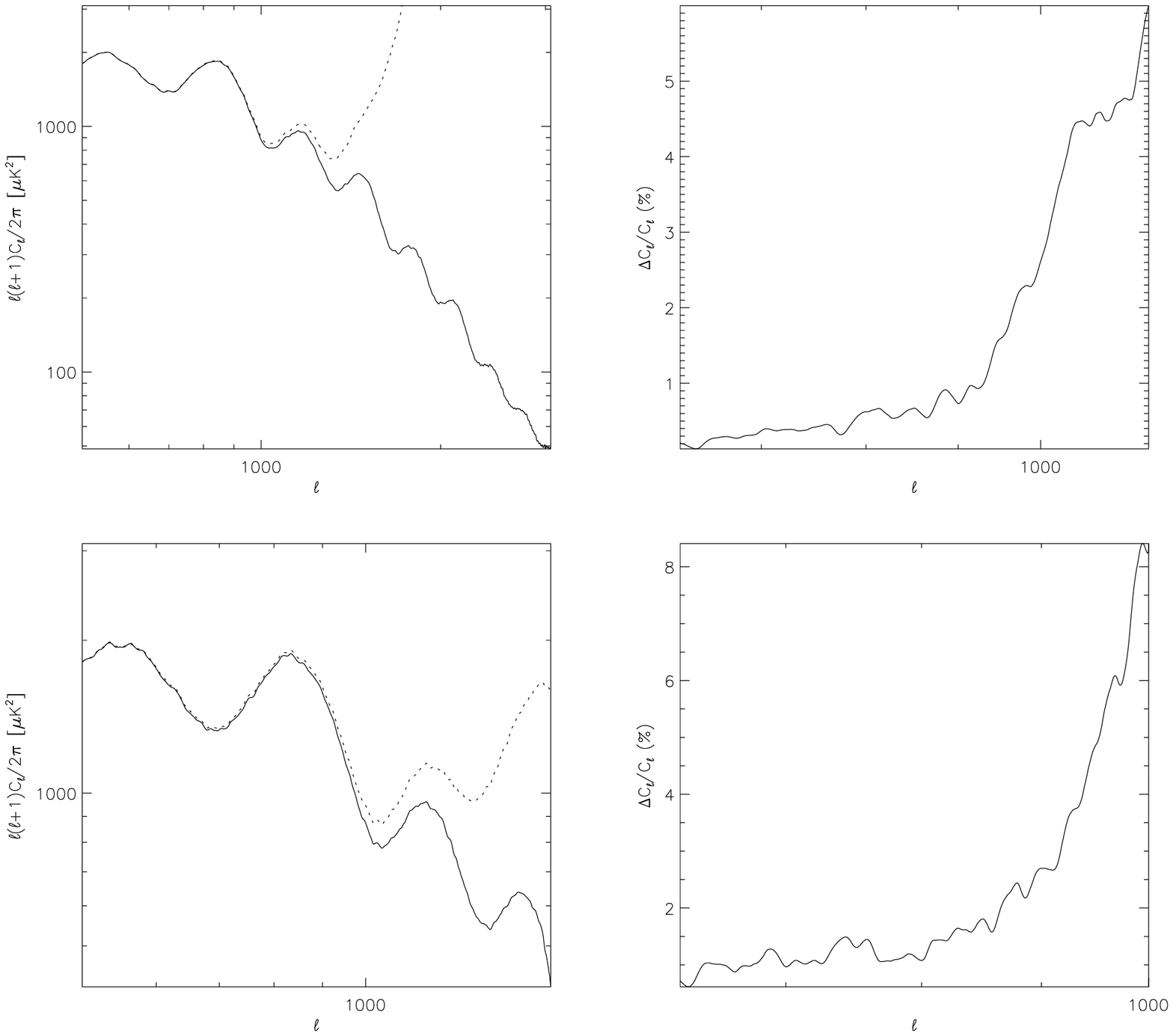}
   \caption{The same as in Fig.~3 but in the presence of
a level of noise $\sigma_{_{N}} = 9 \ \mu K$ for the
pixel size $\Delta = 3.43$ (top panels) and 
$\sigma_{_{N}} = 4.5 \ \mu K$ for
$\Delta = 6.86$ (bottom panels).
No correction for noise effects is performed (see also the text).}
    \end{figure*}

Fortunately, the angular power spectrum of the map
$N^{*}=B^{-1}N$
can be estimated
and subtracted from that of the $T^{*}$ map. In order to do that,
we first solve the matrix equation:
\begin{equation}
N^{\prime} = B N^{*}
\, .
\label{rum}
\end{equation}
This equation can be also solved using the Jacobi method.
The independent terms are the temperatures $N^{\prime}$ corresponding
to a new noise realization different from $N$.
We can now extract the $C_{\ell}$ quantities from the map
$N^{*}$. New numerical errors should appear when we use
the Jacobi method in the presence of noise.
On the other hand, we can try a Monte Carlo approach 
to evaluate the deconvolution effect on the noise.
We can take various $N^{\prime}$ noise realizations
to get an average spectrum of the corresponding $N^{\*}$
maps; in this way, the effect of 
%partial coverage
the noise variance in the estimate of the angular power spectrum of $N^{*}$
is strongly reduced (forty noise realizations suffice).
When we subtract
this spectrum from that of $T^{*}$, namely, when we correct
the $T^{*}$ spectrum taking into account noise effects,
results are much better than those showed in
Fig.~4~\footnote{We observe that an analogous approach can be pursued 
also in the presence of correlated noise, provided that the noise properties 
can be known from laboratory measures and/or directly reconstructed
from the data (Natoli et al. 2002). Of course, in this context, destriping
(or, possibly, map-making, see e.g. Natoli et al. 2001) should be previously
applied both to the data and to the simulated pure noise data.}. 
These new results are presented in Fig.~5. The structure of
this figure is identical to that of Figs.~3 and 4. As in Fig.~3,
the range of $\ell $ values --~in the left panels~--
has been appropriately chosen to
include the region where the displayed curves separate significantly.
The deconvolution is better than that of Fig.~4, where
no correction for the noise has been considered. According to the
right panels of Fig.~5, the relative deviation produced by
deconvolution plus correction is
smaller than
five per cent for $\ell \leq 1500$ ($\ell \leq 1300$)
in the case $\Delta = 3.43$ ($\Delta = 6.86$).
For $\Delta = 3.43$ ($\Delta = 6.86$), deconvolution works very well 
up to the end of the fifth (fourth) acoustic peak.
For equivalent levels of noise in different pixels, we see that 
--~as expected~--
deconvolution has recovered more  
$C_{\ell} $ coefficients for $\Delta = 3.43$; hence, we can say that
results corresponding to  $\Delta = 3.43$ are
sensibly better than those of $\Delta = 6.86$. On account of 
this fact and also for the sake of briefness,
pixels with $\Delta = 6.86$ are not considered
in cases C and D below.

   \begin{figure*}[t]
   \centering
   \includegraphics[width=18cm]{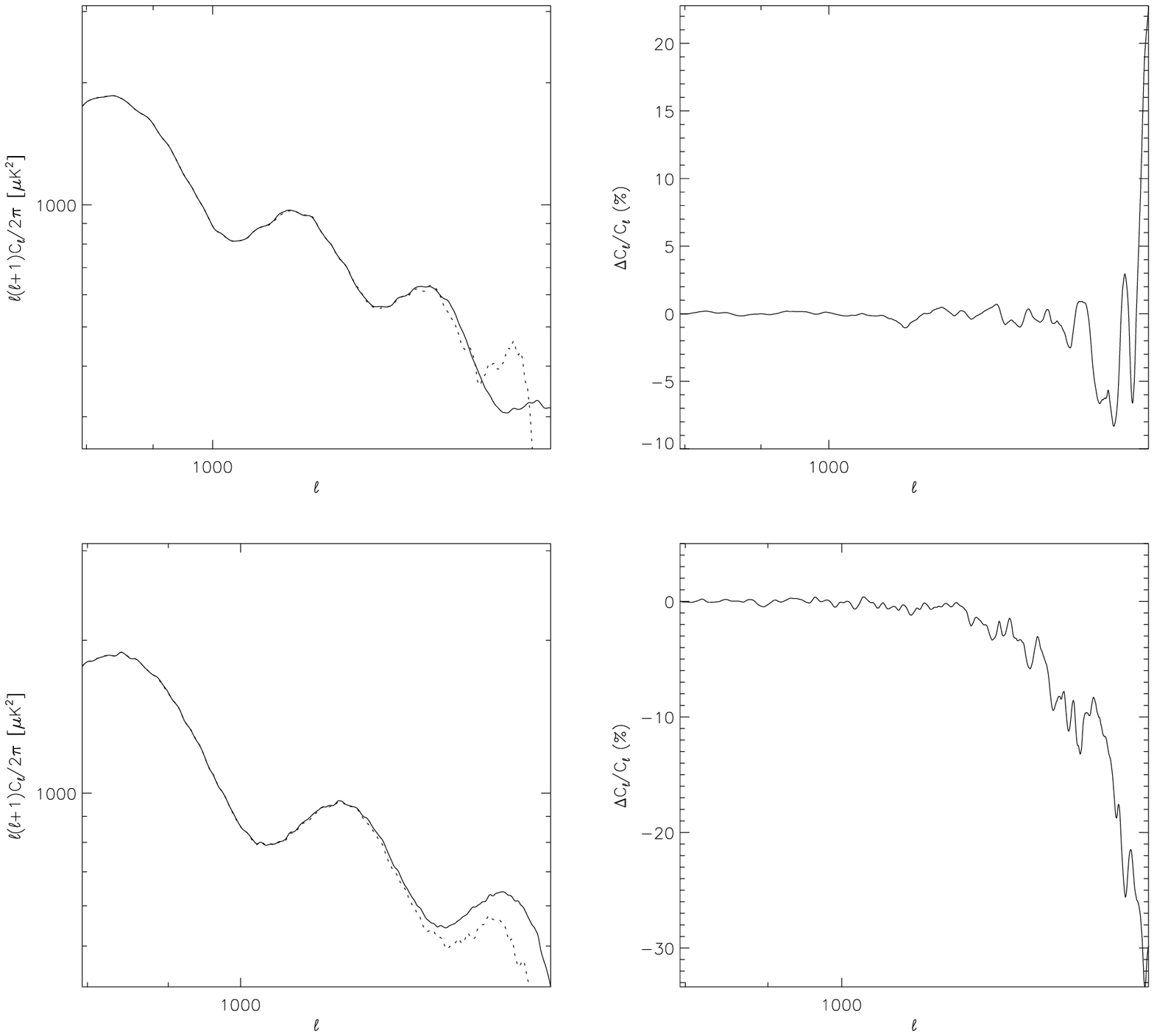}
   \caption{The same as in Fig.~4, but with correction
for noise effects on the angular power spectrum (see also the text).}
    \end{figure*}

\subsection{Case C: application to {\sc Planck} in the absence of noise}

The selected orbit for {\sc Planck} is a Lissajous orbit
around the Lagrangian point L2 of the Sun-Earth system
(see e.g. Bersanelli et al. 1996).
The spacecraft spins at 1 r.p.m.
and the field of view of the two instruments 
--~LFI and HFI (High Frequency Instrument, Puget et al 1998)~-- 
is about $10^\circ \times 10^\circ$ centered at the
telescope optical axis (the so-called telescope line of sight, LOS) 
at a given angle $\alpha$ from the spin-axis direction,
given by a unit vector, $\vec s$, chosen to be pointed 
in the opposite direction with respect to the Sun.
In this work we consider values of  $\alpha \sim 85^{\circ}$,
as adopted for the baseline scanning strategy.
The spin axis will be kept parallel to the Sun--spacecraft direction
and repointed by $\simeq 2.5'$ every $\simeq 1$~hour (baseline
scanning strategy).
Hence {\sc Planck} will trace large circles in the sky
and we assume here, for simplicity, 60 exact repetitions
of the set of the pointing directions of each scan circle.
A precession of the spin-axis with a period, $P$, of $\simeq 6$~months at
a given angle $\beta \sim 10^{\circ}$
about an axis, $\vec f$, parallel to the Sun--spacecraft direction
(and outward the Sun) and shifted of $\simeq 2.5'$ every $\simeq 1$~hour,
may be included in the scanning strategy, possibly with 
a modulation of the speed of the precession in order to optimize
data transmission (Bernard et al. 2002). 
The quality of our deconvolution code is of course
almost independent of the details
of these proposed scanning strategies, and we assume here the 
baseline scanning strategy for sake of simplicity.

The code implemented for simulating
{\sc Planck} observations for a wide set of scanning
strategies is described in detail in Burigana et al. (1997, 1998)
and in Maino et al. (1999).
In this study we do not include the effects introduced 
by the {\sc Planck} orbit, to be currently optimized,
by simply assuming {\sc Planck} located in L2, because
they are fully negligible in this context.

We compute the convolutions between the antenna pattern response
and the sky signal as described in Burigana et al. (2001)
by working at $\sim 3.43'$ resolution 
and by considering spin-axis shifts of $\sim 2.5'$ every hour
and 7200 samplings per scan circle. 
We simulate 11000 hours
of observations (about 15 months) necessary to complete two 
sky surveys with the all {\sc Planck} beams.

With respect to the reference frames described in 
Burigana et al. (2001), following the recent developments 
in optimizing the polarization properties of LFI main beams
(see e.g. Sandri et al. 2003), the conversion between the 
standard Cartesian {\it telescope frame} $x,y,z$ and the 
{\it beam frame} $x_{bf},y_{bf},z_{bf}$ actually requires 
a further angle $\psi_B$ other than
the standard polar coordinates $\theta_B$ and $\phi_B$
defining the colatitude and the longitude of the main beam 
centre direction in the {\it telescope frame}.
Appendix~A provides the transformation rules 
between the {\it telescope frame} and the {\it beam frame}, as well
as the definition of the reference frames adopted in this work. 

The orientation of these frames as the satellite moves is implemented in
the code. For each integration time, we determine the orientations
in the sky of the {\it telescope frame} and of the {\it beam frame},
thus performing a direct convolution with the sky signal
by exploiting the detailed main beam response
in each considered sky direction.
The detailed main beam shape and position on the telescope field of view
adopted in this application is that 
computed in the past year for the feed LFI9 (Sandri et al. 2002)
which shows an effective FWHM resolution of $10.68'$ and deviations 
from the symmetry producing a typical ellipticity ratio of 1.25.
Such values of resolution and asymmetry parameter are in the range 
of them that it is possible to reach with a 1.5~m telescope 
like that of {\sc Planck} by optimizing the optical design 
(see e.g. Sandri et al. 2003). Although our deconvolution method
is largely independent of the details of the considered beam shape, it is interesting
to exploit its reliability under quite realistic conditions.

The CMB anisotropy map has been projected into the HEALPix scheme
%({\it Hierarchical Equal Area and IsoLatitude Pixelization
%of the Sphere}) 
(G\`orski et al. 1999) starting from the angular
power spectrum of the assumed $\Lambda$CDM model
(see Sect.~1).

%, $C_\ell$, computed with the CMBFAST code 
%by Seljak \& Zaldarriaga (1996) for a flat $\Lambda$CDM model. 

To make the application of the deconvolution code easier
and the system solution possible 
without large RAM requirement and in a reasonable computational 
time~\footnote{In the current implementation, 
about 19 hours of computation are required to deconvolve a single 
patch with $1024^2$ pixels on an 64~bit alpha digital unix machine with 
single cpu at 533~MHz and 1~Gb RAM.}
we implemented a code that identify in the simulated 
time ordered data (TOD) all and only the beam centre pointing directions
in an equatorial patch (in ecliptic coordinates) 
of $1024 \times 1024$ pixels with a $\sim 3.43'$ side
($n_{side}=1024$). 
We keep the exact information on the beam centre pointing direction 
and the beam orientation (defined for instance by an angle
between the axis $x_{bf}$ and the parallel in the beam centre pointing
direction) as computed by our flight simulator. All the samples 
of the TOD within the same pixel are identified and restored 
in contiguous positions. 
At this aim, we take advantage from the 
nested, hierarchical ordering of the HEALPix.
This is quite simple
in the current simplified simulation, but it will require 
the development of efficient and versatile tools to manage 
the more general case
in which the all samples from the experiment multi-beam array  
are considered, particularly for the ecliptic polar patches, which
pixels are observed many and many times because of the {\sc Planck}
scanning strategy.
In the context of the {\sc Planck} project, 
this effort will be pursued by taking advantage from the 
development of {\sc Planck} Data Model 
(see e.g. Lama et al. 2003).

>From the simulated TOD, possibly restored as described above, 
we extract a map of a patch of simply coadded data and a map 
of a patch deconvolved by applying method II. The latter map
can be then symmetrically smoothed with a beam FWHM of $10.68'$ 
by using the HEALPix tools
for comparison with the former one, obtained from the 
convolution with the simulated asymmetric 
beam and taking into account the scanning strategy. 
Of course, from the input map we can extract the same sky patch.

We consider four different patches covering 
an equatorial region of $\simeq 28.3$~\% of the sky 
(analogously to the case of small patches, see Sect.~3.1,
avoiding the boundary regions of the four patches slightly reduces 
the originally considered, $\simeq 33.3$~\%, sky coverage).

All the above maps are inverted with the {\it anafast} code
of HEALPix to extract the correponding angular power spectra.
The result is shown in Fig.~6. 
Of course, all the angular power spectra are 
in strict agreement at multipoles $\lsim 200$, where
the main beam distortion effect is negligible for 
a beam with a FWHM of about $10'$ and a reasonable ellipticity.
Note the very good agreement 
between the power spectrum of the input map and that derived with
method II: the difference becomes significant only 
at the seventh acoustic peak
[compare the solid (black) line with the dashed (green) line].
Note the power excess at high $\ell$
introduced by the beam distortions, compared with the  
power spectrum derived from the deconvolved map subsequently 
symmetrically smoothed [compare the dash-three dots (fuxia)
with the dash-dots (blue)].
Also, the power spectrum derived from coadded map 
when divided by the window function corresponding to
the symmetric equivalent beam, ${\rm exp}[-(\sigma\ell)^2]$,
significantly exceeds that of the input map at $\ell$ larger
than the fourth acoustic peak [compare the dots (red)
with the solid line (black)]. We find a similar disagreement
even by varying the assumed value of the symmetric beam 
width $\sigma$: an improvement on a limited range of multipoles 
results in a worsening on a different range of 
multipoles.

This demonstrates 
that a kind of deconvolution is necessary to remove the 
main beam distortion effect at very high multipoles
and that method II works well in the absence of noise.
The impact of instrumental noise is discussed in the next subsection.

   \begin{figure*}[t]
   \centering
   \begin{tabular}{cc}
   \includegraphics[width=9cm]{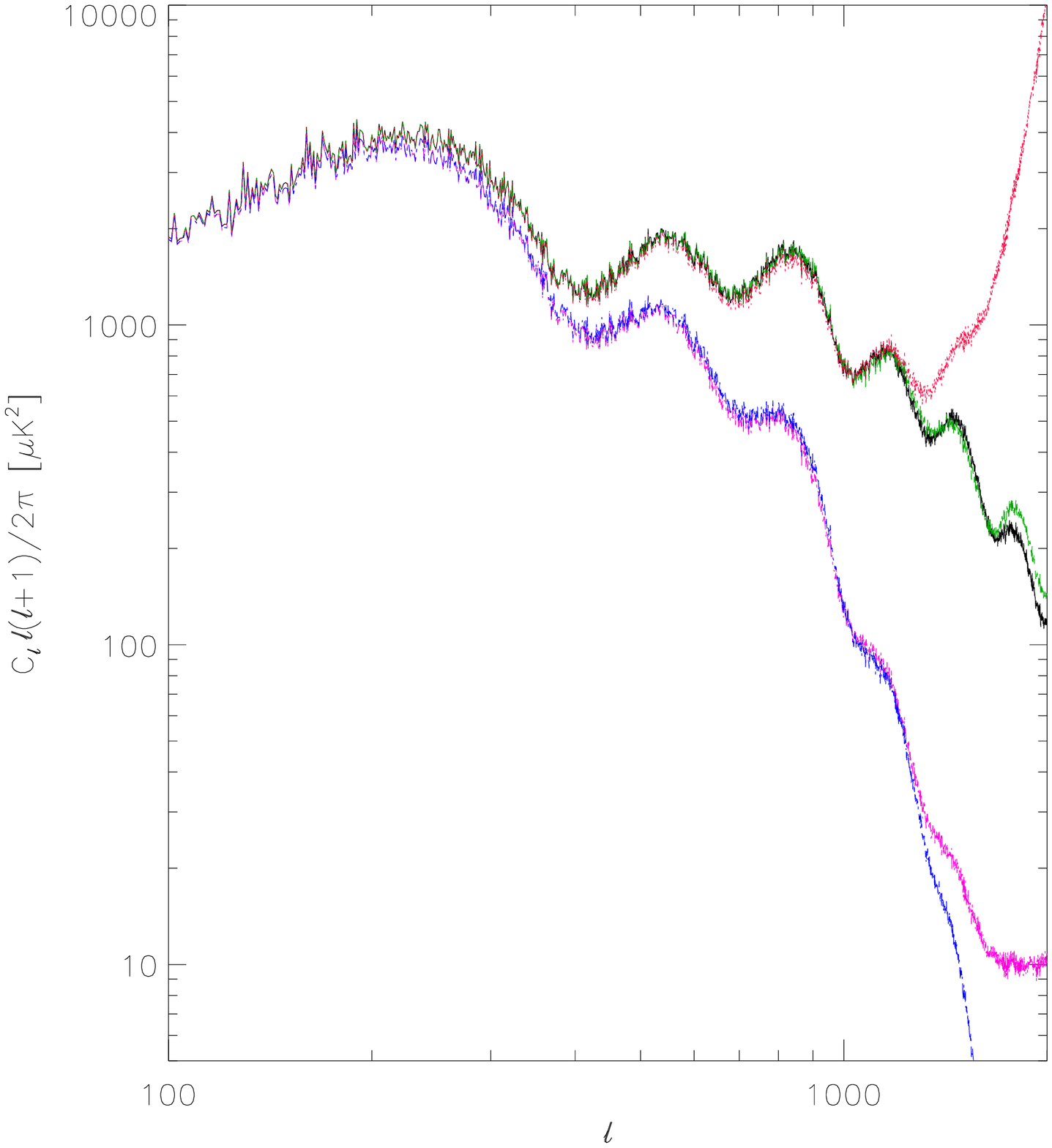}
   \includegraphics[width=9cm]{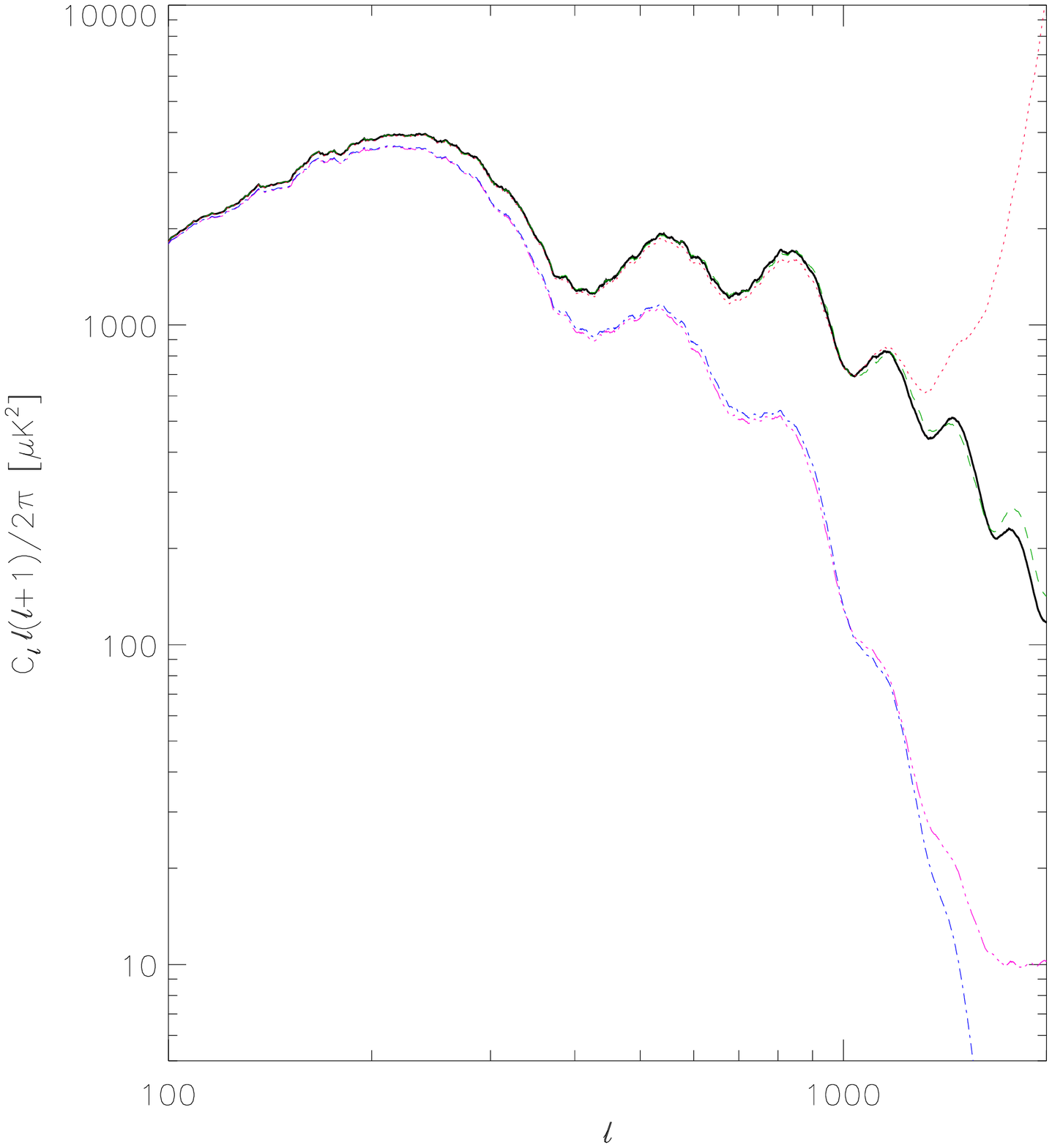}
   \end{tabular}
   \caption{The two panels are identical except for the binning
in $\ell$ in the right panel. No noise is considered. 
Angular power spectrum from the four considered patches.
Solid (black) line: angular power spectrum of the input map 
without convolution; 
dash-three dots (fuxia): angular power spectrum of the map convolved 
with the simulated beam assuming the {\sc Planck} scanning
strategy; dots (red): angular power spectrum of the map convolved 
with the simulated beam assuming the {\sc Planck} scanning
strategy and divided by the symmetric beam window function;
dashes (green): angular power spectrum of the deconvolved map;
dash-dots (blue): angular power spectrum of the deconvolved map subsequently
convolved with a symmetric beam with the effective resolution
(see also the text).}
    \end{figure*}

\subsection{Case D: application to {\sc Planck} in the presence of noise}

Analogously to the Case~B (see Sect.~3.2) we have 
simulated four patches (of $1024^2$ pixels, as in the previous section)  
of instrumental uncorrelated Gaussian noise
with $\sigma_N = 9 \mu$K (in terms of antenna temperature) 
for a pixel of $3.43'$,
as appropriate to the global {\sc Planck} sensitivity at 100~GHz;
for simplicity we have assumed a uniform noise. 

This realization of noise map has been added to the map of signal 
and then method~II has been applied to deconvolve
the map of signal plus noise, as described in Sect.~3.2.
We produced also four realizations of pure noise maps 
to be deconvolved in the same way. Finally, we generated
another map of noise to be superimposed to the
coadded map obtained from the convolution with the simulated beam
including scanning strategy, for comparison.

We computed the angular power spectrum of the four maps of pure noise 
and of the four maps of pure noise deconvolved with method II.
Fig.~7 (left panel) compares the averages of 
the four realizations of these power spectra and their relative
variance (right panel). As evident, deconvolution increases the noise:
a rough approximation of the ratio between of the noise angular power spectrum 
after deconvolution and before deconvolution 
is given by 
$\sim 2({\rm FWHM}/\Delta)^2 {\rm exp}[(\sigma \ell/2)^2+(\sigma \ell/2)^6]$
where, as usual, ${\rm FWHM}=\sqrt{8{\rm ln}2} \sigma$ ($=10.68'$) 
and $\Delta$ is the pixel side ($=3.43'$).
On the other hand, the relative variance of these power spectra
is almost similar.

   \begin{figure*}[t]
   \centering
   \begin{tabular}{cc}
   \includegraphics[width=9cm]{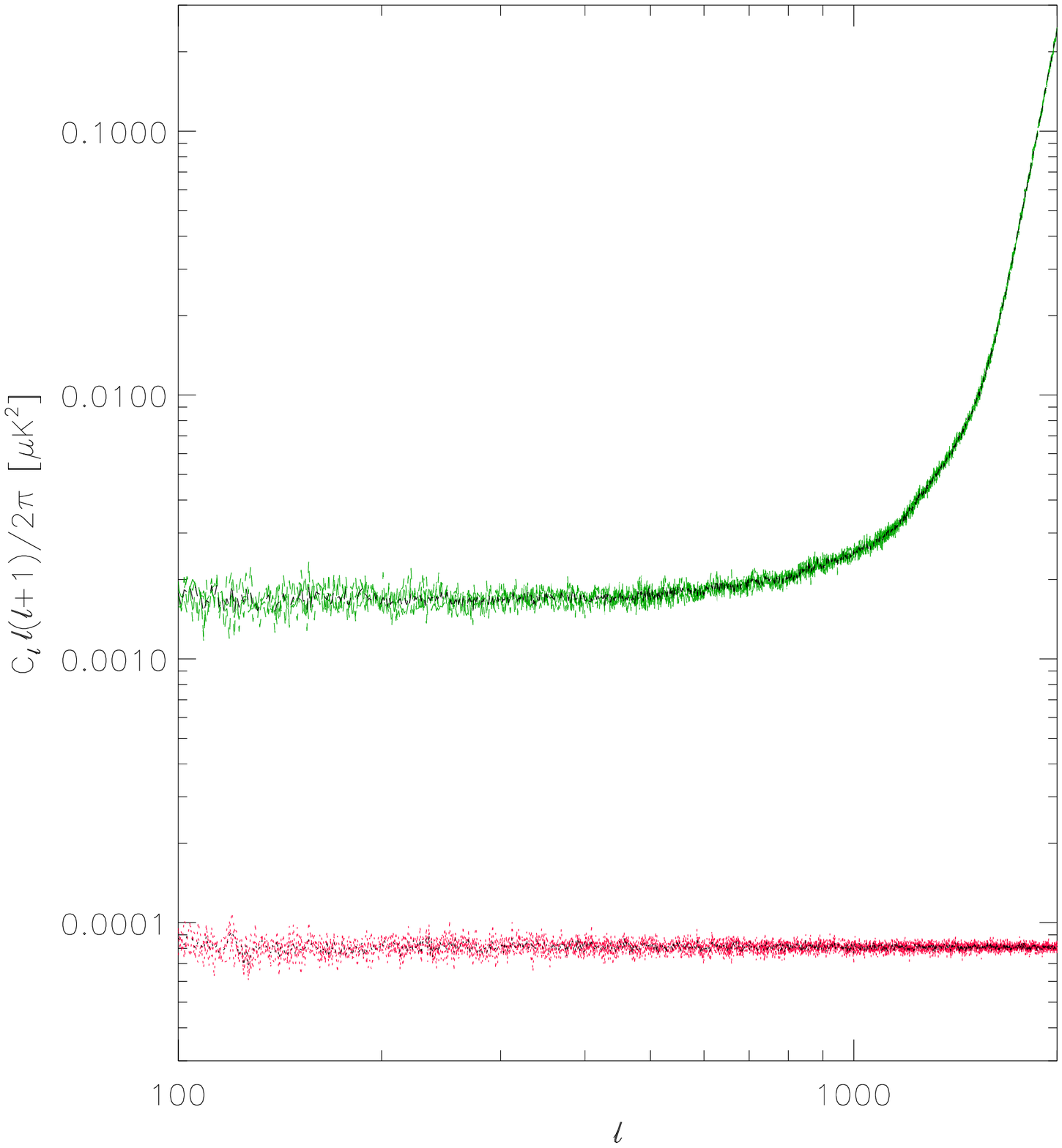}
   \includegraphics[width=9cm]{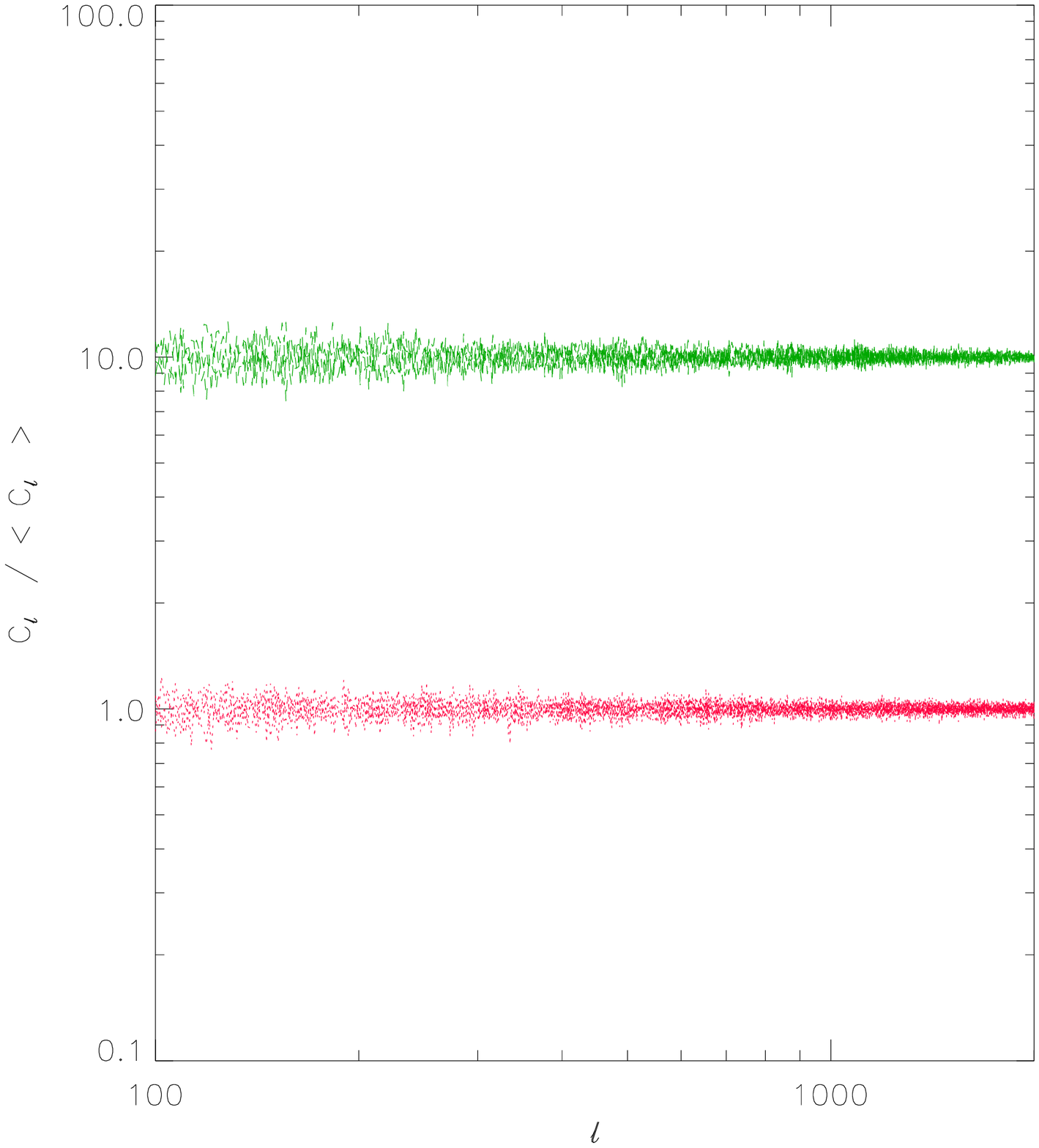}
   \end{tabular}
   \caption{Left panel: comparison between the noise power spectrum before (red)
and after (green) deconvolution for the four noise realizations 
(the inner (black) ``curves'' represent the two average noise power spectra). 
Right panel: ratio between the power spectrum of each of the four 
noise realizations and the average noise power spectrum, 
before and after deconvolution, multiplied by 10 in the latter  
case for graphic purposes (see also the text).}
    \end{figure*}

In spite of the relatively large increase of the noise power, 
we find that method II
results to work quite well in removing the effect of main beam 
distortions up to the end of the fifth acoustic peak, 
when the average power spectrum  
of the deconvolved pure noise maps is subtracted to the 
power spectrum of the deconvolved noisy map
(see the dashed (green) line in Fig.~8). 

   \begin{figure*}[t]
   \centering
   \begin{tabular}{cc}
   \includegraphics[width=9cm]{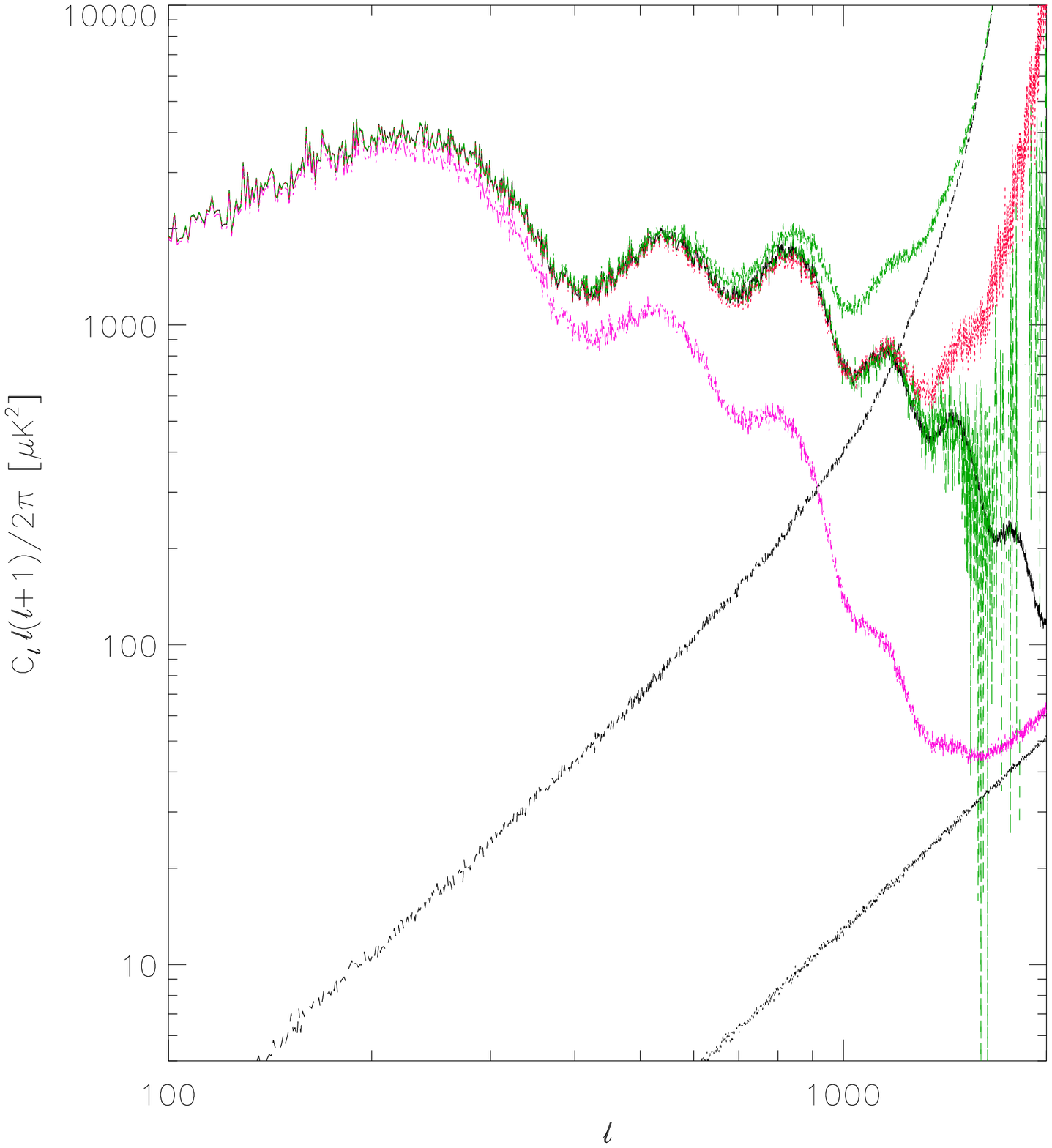}
   \includegraphics[width=9cm]{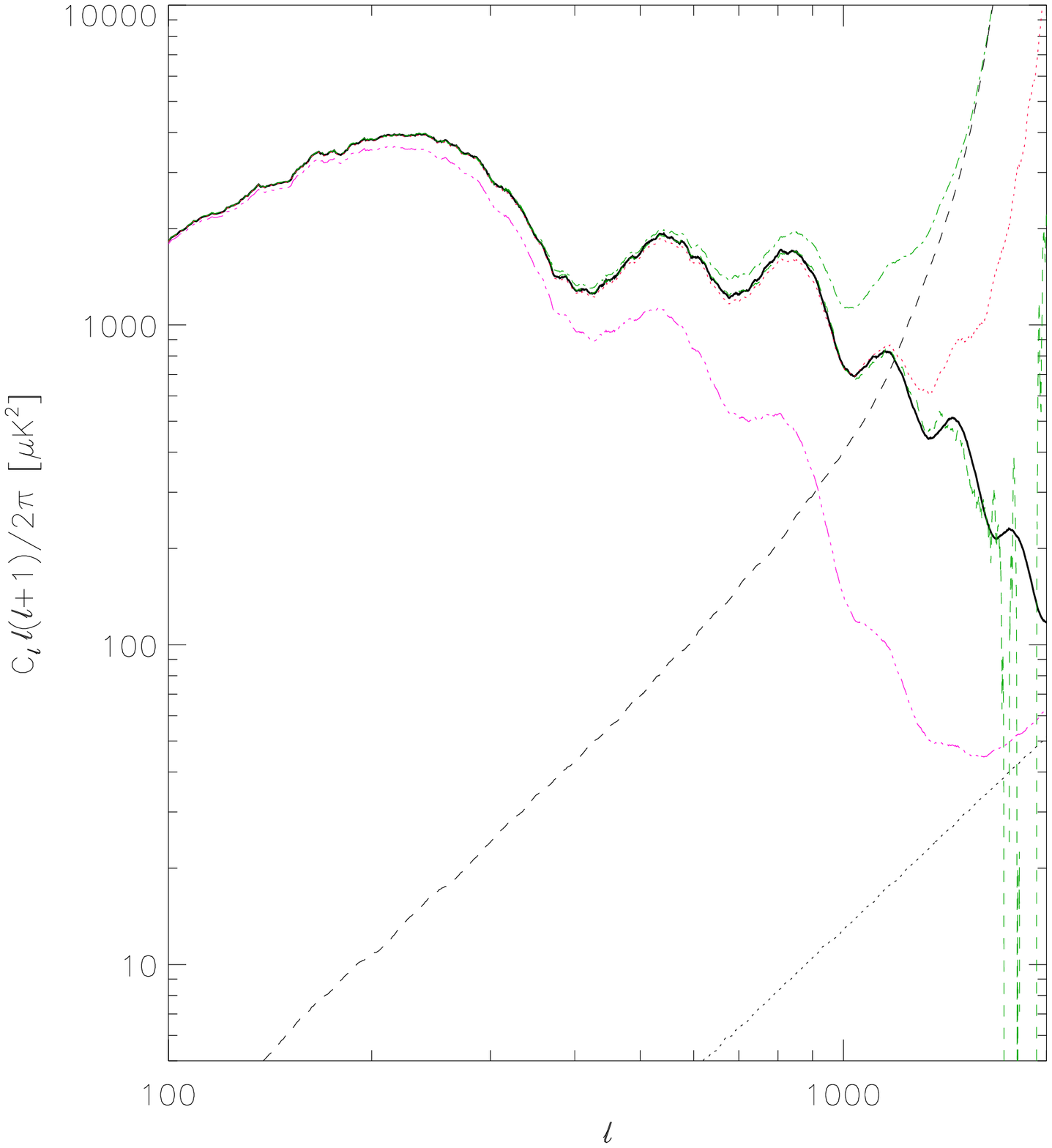}
   \end{tabular}
   \caption{The two panels are identical except for the binning
in $\ell$ in the right panel. The noise is included.
Angular power spectrum from the four considered patches.
Solid (black) line: angular power spectrum of the input map 
without convolution and without noise; 
dash-three dots (fuxia): angular power spectrum of the map convolved 
with the simulated beam assuming the {\sc Planck} scanning
strategy and adding a noise realization; 
dots (red): angular power spectrum of the map convolved 
with the simulated beam assuming the {\sc Planck} scanning
strategy and adding a noise realization, 
after the subtraction of the averaged power spectrum 
of four noise realizations and then divided by the symmetric 
beam window function;
dotted (black) bottom line: averaged power spectrum
of four noise realizations; 
dash-dots (green): angular power spectrum of the deconvolved map
in the presence of a noise realization;
dashes (green): angular power spectrum of the deconvolved map 
in the presence of a noise realization
after the subtraction of the averaged power spectrum
of four noise realizations deconvolved in the same way;
dashed (black) bottom (at low $\ell$) line: averaged power spectrum
of four noise realizations deconvolved in the same way
(see also the text).}
    \end{figure*}

In Fig.~9 we report the relative (per cent) errors 
introduced by beam distortions in the absence of deconvolution, 
in the presence of deconvolution without applying the subtraction of the 
average deconvolved noise spectrum 
and by applying the deconvolution and the subtraction of average deconvolved
noise spectrum. As evident, in the last case the power spectrum can be recovered 
with a good accuracy up to high multipoles
(relative errors $\lsim 5, 10, 15, 20$~\% respectively for
$\ell \lsim 1250, 1470, 1500, 1650$ --~see the dashed (green) line in the middle panel~-- 
to be compared with errors $\simeq 10, 20, 30$~\% at $\ell \simeq 1200, 1250, 1300$ 
--~see the dotted (red) line in the middle panel~-- and then 
dramatically increasing with $\ell$ in the absence of deconvolution). 
The right panel shows 
what already found for the noiseless case (Sect.~3.3): 
even by varying the assumed value of the symmetric beam 
width $\sigma$, an improvement in the $C_\ell$ recovery can not be
reached simultaneously on the whole relevant range of multipoles.

   \begin{figure*}[t]
   \centering
   \begin{tabular}{ccc}
   \includegraphics[width=6cm]{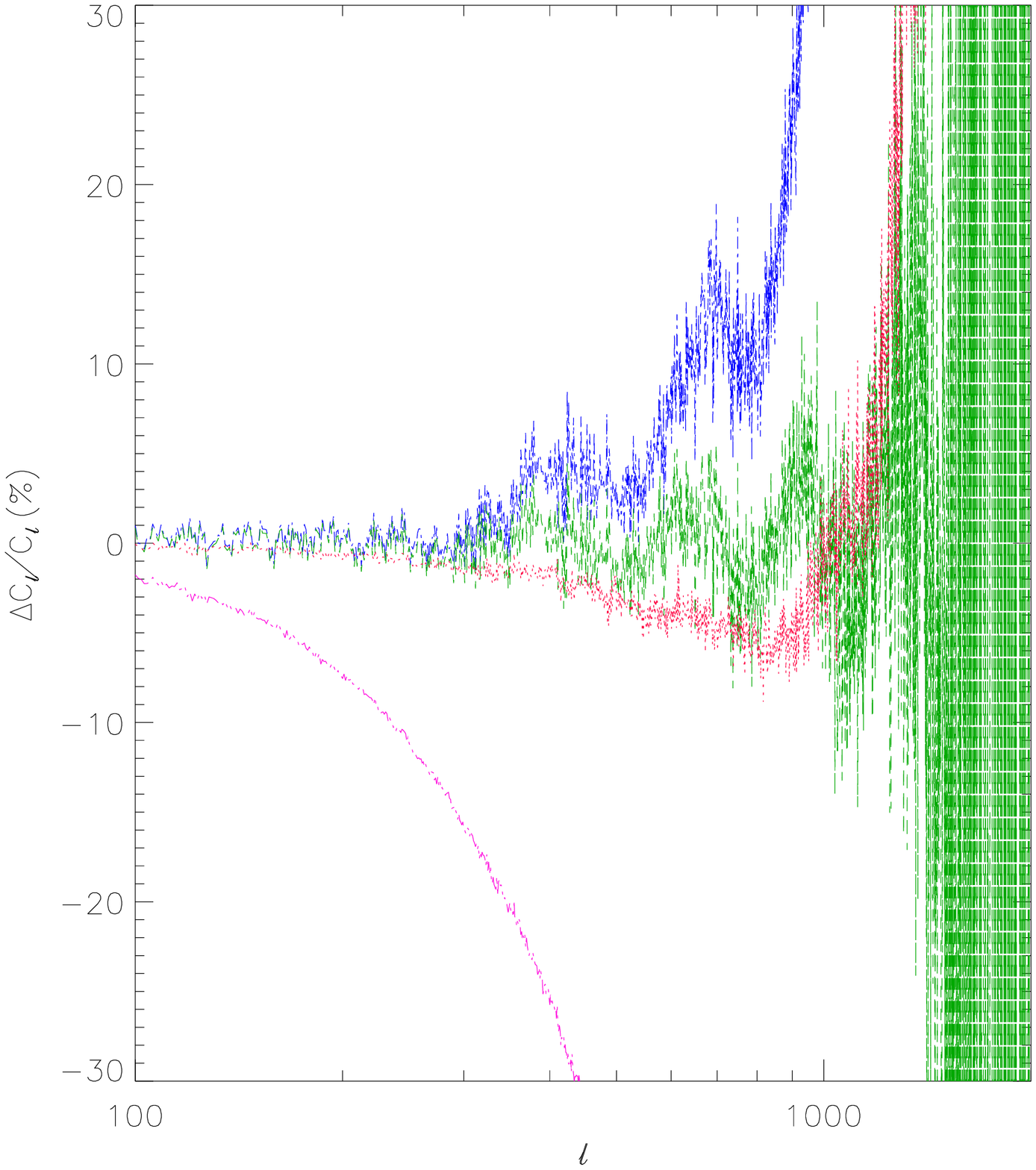}
   \includegraphics[width=6cm]{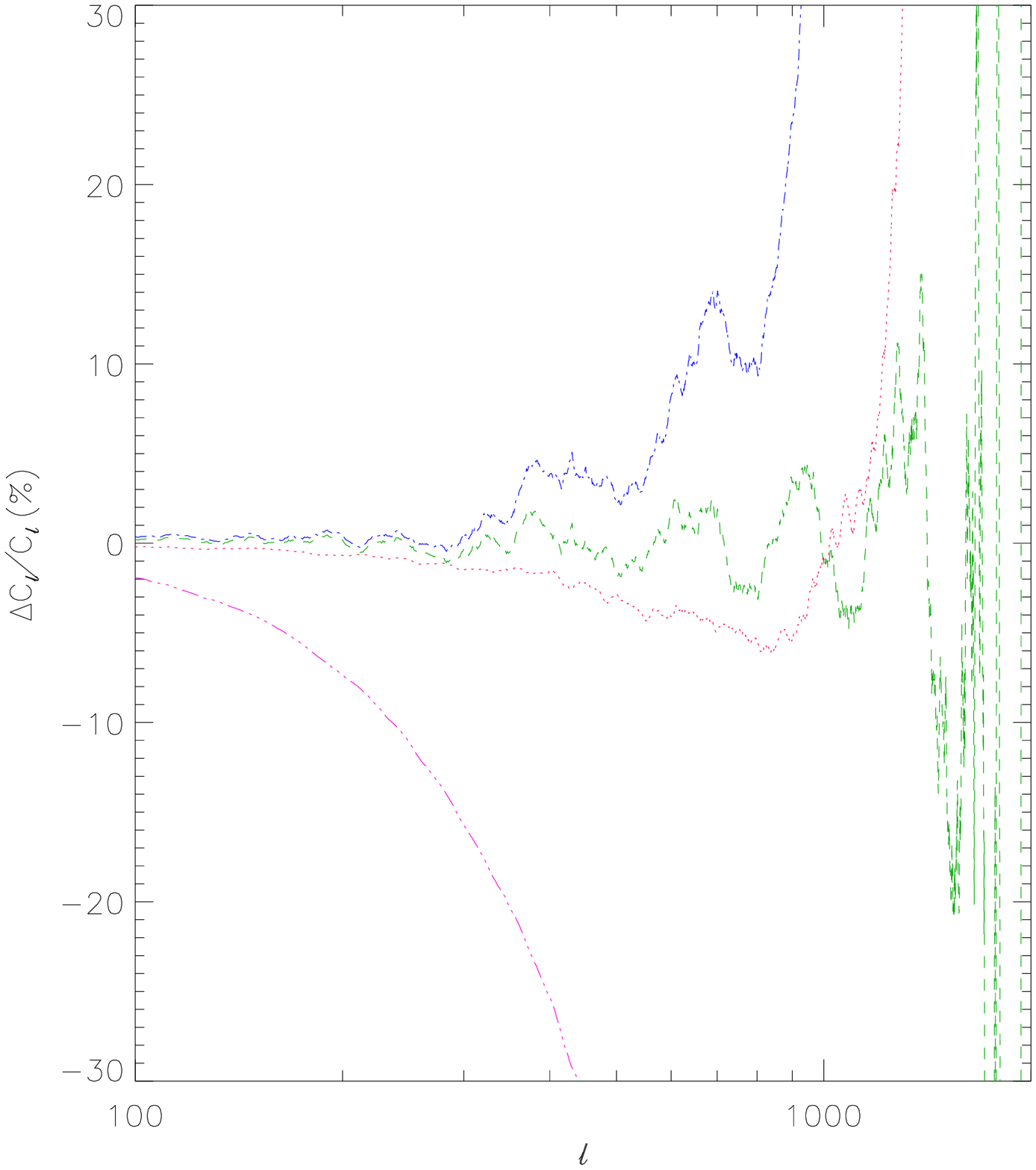}
   \includegraphics[width=6cm]{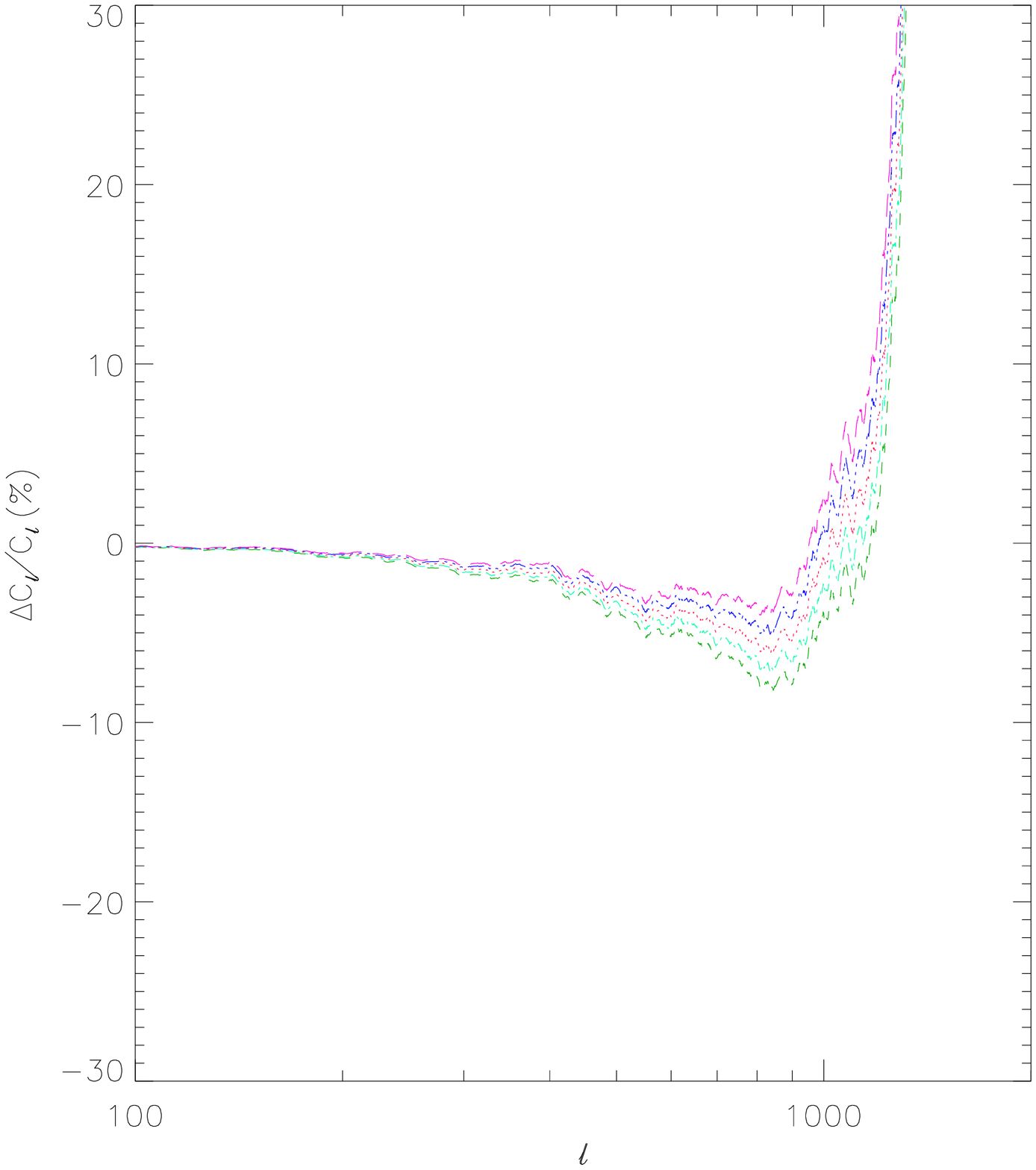}
   \end{tabular}
   \caption{The left and middle panels are identical except for the binning
in $\ell$ in the middle panel. 
Errors in the angular power spectrum recovery from the four considered patches.
Dash-three dots (fuxia): angular power spectrum of the map convolved 
with the simulated beam assuming the {\sc Planck} scanning
strategy and adding a noise realization; 
dots (red): angular power spectrum of the map convolved 
with the simulated beam assuming the {\sc Planck} scanning
strategy and adding a noise realization, 
after the subtraction of the averaged power spectrum 
of four noise realizations and then divided by the symmetric 
beam window function;
dash-dots (blue): angular power spectrum of the deconvolved map
in the presence of a noise realization;
dashes (green): angular power spectrum of the deconvolved map 
in the presence of a noise realization
after the subtraction of the averaged power spectrum
of four noise realizations deconvolved in the same way.
In the right panel we report the same line (dots, red)
of the middle panel and four other power spectra
obtained in similar way by assuming different values
of effective angular resolution, with an effective FWHM increased
or decreased by $0.05'$ and $0.1'$. Clearly, an improvement
at $\ell \sim 400-800$ implies a worsening at $\ell \gsim 1000$ 
and viceversa: therefore, even allowing for
changes in the assumed effective angular resolution, 
the simple symmetric beam approximation
can not improve the power spectrum recovery 
simultaneously in the two above ranges of $\ell$ (see also the text).}
    \end{figure*}

We have considered here for simplicity only equatorial patches.
On the other hand, we have verified
that the main beam distortion affects 
the reconstructed power spectrum in a similar way 
also on polar patches.
In fact, even at high ecliptic latitudes 
a given pixel is preferentially observed with a limited
number of beam orientations.
In polar patches the {\sc Planck} sensitivity is significantly 
better than the average. 
Therefore, in spite of the more complex data storage,
deconvolution will be there less affected by the 
noise~\footnote{For example, the average sensitivity 
on a polar region of about 25 squared degrees
is about 5 times better (i.e. $\simeq 2 \mu$K on pixel
of $3.43'$!) than the average full sky sensitivity:
then, we expect there a deconvolution quality intermediate between 
that found here and that found in the previous section.}.

\section{Discussion and conclusions}

We have presented the basic formalism for a robust and 
feasible method for deconvolution in noisy CMB maps. 

We have implemented and tested this method for two completely different
situations: small patches observed with an elliptical beam and 
with a scanning strategy 
involving repeated measures of the same pixel but
not related to a specific experiment and quite large sky areas 
observed with a realistic beam and a scanning strategy 
like that of the {\sc Planck} satellite.
A sensitivity level and a beam resolution 
typical of the {\sc Planck} experiment at 100~GHz have been exploited.

After having verified the good accuracy of our deconvolution code
to remove the main beam distortion effect in the angular power 
spectrum estimation in the absence of noise, we applied it to 
noisy maps. We demonstrate that it is possible to accurately evaluate
the effect of deconvolution on pure noise simulated maps
so deriving, with Monte Carlo simulations, a good estimation
of the average deconvolved noise angular power spectrum to be subtracted
from the deconvolved noisy maps.

In practice, for the considered sensitivity, $\simeq 9\mu$K 
for a pixel of $3.43'$, and beam resolution, FWHM~$\simeq 10'-11'$, 
our deconvolution code allows to efficiently 
remove the main beam distortion effect and accurately reconstruct
the CMB power spectrum up to the end of the fifth acoustic peak. 

Clearly, in the context of the {\sc Planck} project,
the measure of the very high multipole region of the 
CMB angular power spectrum will be take advantage from the
cosmological frequency channels at highest resolution, 
namely the 217~GHz channel (having a FWHM~$\simeq 5'$), 
where Poisson fluctuations from
extragalactic sources (see e.g. Toffolatti et al. 1998) 
are expected to be at a very low level and anisotropies
from thermal Sunyaev-Zeldovich effects are, if not 
exactly null because of possible unbalanced contributions 
within the bandwidth, certainly very small.
On the other hand, the frequency range about 100~GHz is where
the global (Galactic plus extragalactic) foreground contamination
is expected to be minimum. Therefore, it is extremely relevant
to extract at these frequency an accurate estimation of the sky
angular power spectrum, cleaned, as better as possible,
from the all systematic effects. 
In addition, the removal of the main beam distortion effect,
relevant at large multipoles, greatly helps the comparison between the
results obtained at different frequency channels.

Of course, we plan to apply this method in the future 
also to lowest and highest beam resolutions and in the presence 
of other kinds of systematic effects. 
We believe that the results found here are very encouraging, suggesting that
the main beam distortion effect, previously reduced by optimizing 
the optical design, can be further reduced in the data analysis.

\begin{acknowledgements}
This work has been partially
supported by the Spanish MCyT
(project AYA2000-2045).
A part of the calculations
was carried out on a SGI Origin 2000s at the Centro de Inform\'atica
de la Universidad de Valencia. It is a pleasure to thank the LFI DPC, 
were some computations were carried out, 
and the staff working there for many useful discussions.
We warmly thank M.~Sandri and F.~Villa for having provided us
with one of the LFI main beam, simulated at high resolution, 
adopted in some parts of this work.
Some of the results in this paper have been derived using the HEALPix
(G\`orski et al. 1999).
We thank U. Seljak and M. Zaldarriaga for the use of 
the CMBFAST code. 
%We wish to thank the referee for constructive comments.
\end{acknowledgements}

\appendix

\section{Transformation rules between {\sc Planck} {\it telescope frame} 
and {\it beam frame}}

Let $\vec s$ be the unit vector, choosen outward the Sun direction,
of the spin axis direction and 
$\hat{k}$ that of the direction, $z$, of the telescope line of sight (LOS),
pointing at an angle $\alpha \sim 85^{\circ}$ from the direction of $\vec s$. 

On the plane tangent to the celestial 
sphere in the direction of the LOS
we choose two coordinates $x$ and $y$, respectively defined by the unit vector
$\hat{i}$ and $\hat{j}$
according to the convention that the unit vector
$\hat{i}$ points always toward $\vec s$
and that $x,y,z$ is a standard Cartesian frame,
referred here as {\it telescope frame}.

Let $\hat{i}_{bf},\hat{j}_{bf},\hat{k}_{bf}$ be
the unit vectors corresponding to the Cartesian axes 
$x_{bf},y_{bf},z_{bf}$ of the {\it beam frame}; $\hat{k}_{bf}$ defines the  
direction of the beam centre axis in the {\it telescope frame}.
The {\it beam frame} is defined with respect to the {\it telescope frame}
by three angles: $\theta_B$,~$\phi_B$,~$\psi_B$ ($\theta_B$ and $\phi_B$,
two standard polar coordinates defining the direction of the 
beam centre axis, range respectively from $0^\circ$, for an on-axis beam, to some
degrees, for LFI off-axis beams, and from $0^\circ$ to $360^\circ$).

Let $\hat{i'}_{bf},\hat{j'}_{bf},\hat{k'}_{bf'}$ 
($\hat{k'}_{bf}=\hat{k}_{bf}$) be the unit vectors corresponding to 
the Cartesian axes $x',y',z'$ of an {\it intermediate frame}, defined
by the two angles $\theta_B$ and $\phi_B$, obtained by
the {\it telescope frame} $x,y,z$ when the unit vector of the axis $z$ 
is rotated by an angle $\theta_B$ on the plane defined by the unit vector 
of the axis $z$ and the unit vector $\hat{k}_{bf}$ up to reach $\hat{k}_{bf}$:

%        beamvec(0)=cos(phibeamrad)*sin(thbeamrad)
%        beamvec(1)=sin(phibeamrad)*sin(thbeamrad)
%        beamvec(2)=cos(thbeamrad)
\begin{equation}
\hat{k'}_{bf} = \hat{k}_{bf} =  {\rm cos}(\phi_B) {\rm sin}(\theta_B) \hat{i}
              + {\rm sin}(\phi_B) {\rm sin}(\theta_B) \hat{j} 
              + {\rm cos}(\theta_B) \hat{k}
\end{equation}

\begin{equation}
\hat{i'}_{bf} =  [{\rm cos}(\phi_B)^2 {\rm cos}(\theta_B) + {\rm sin}(\phi_B)^2] \hat{i}
              + [{\rm sin}(\phi_B) {\rm cos}(\phi_B) ({\rm cos}(\theta_B)-1)] \hat{j}
              - {\rm sin}(\theta_B) {\rm cos}(\phi_B) \hat{k} 
\end{equation}

\begin{equation}
\hat{j'}_{bf} =  [{\rm sin}(\phi_B) {\rm cos}(\phi_B) ({\rm cos}(\theta_B)-1)] \hat{i}
              + [{\rm cos}(\theta_B) {\rm sin}(\phi_B)^2 + {\rm cos}(\phi_B)^2] \hat{j}
              - {\rm sin}(\theta_B) {\rm sin}(\phi_B) \hat{k} \, . 
\end{equation}

%        i_mb_inframe_pm(0)=cos(phibeamrad)**2*cos(thbeamrad)
%     &                     +sin(phibeamrad)**2
%        i_mb_inframe_pm(1)=sin(phibeamrad)*cos(phibeamrad)
%     &                     *(cos(thbeamrad)-1.d+0)
%        i_mb_inframe_pm(2)=-sin(thbeamrad)*cos(phibeamrad)
%
%        j_mb_inframe_pm(0)=i_mb_inframe_pm(1)
%        j_mb_inframe_pm(1)=cos(thbeamrad)*sin(phibeamrad)**2
%     &                     +cos(phibeamrad)**2
%        j_mb_inframe_pm(2)=-sin(thbeamrad)*sin(phibeamrad)

The {\it beam frame} is obtained from the {\it intermediate frame}
through a further (anti-clockwise) rotation of an angle $\psi_B$ 
(ranging from $0^\circ$ to 
$360^\circ$~\footnote{We note that, in other conventions, angles $\phi'_B$ and $\psi'_B$
ranging from $-180^\circ$ to $180^\circ$ are given, instead of $\phi_B$ and $\psi_B$. 
The angles $\phi_B$ and $\psi_B$ here defined are 
equal to $\phi'_B$ and $\psi'_B$ when they are positive 
and are given respectively by 
$360^\circ + \phi'_B$ and $360^\circ + \psi'_B$ for negative $\phi'_B$ and $\psi'_B$.})
around $\hat{k}_{bf}$ and is therefore
explicitely given by:

\begin{eqnarray}
\hat{i}_{bf} & = &  [{\rm cos}(\psi_B) \hat{i'}_{bf,x} + {\rm sin}(\psi_B) \hat{j'}_{bf,x}] \hat{i}   
              +  [{\rm cos}(\psi_B) \hat{i'}_{bf,y} + {\rm sin}(\psi_B) \hat{j'}_{bf,y}] \hat{j} \nonumber \\
             & ~ &  +  [{\rm cos}(\psi_B) \hat{i'}_{bf,z} + {\rm sin}(\psi_B) \hat{j'}_{bf,z}] \hat{z}
\end{eqnarray}

%        ii_mb_inframe_pm(0)=cos(psibeamrad)*i_mb_inframe_pm(0)
%     &                      +sin(psibeamrad)*j_mb_inframe_pm(0)
%        ii_mb_inframe_pm(1)=cos(psibeamrad)*i_mb_inframe_pm(1)
%     &                      +sin(psibeamrad)*j_mb_inframe_pm(1)
%        ii_mb_inframe_pm(2)=cos(psibeamrad)*i_mb_inframe_pm(2)
%     &                      +sin(psibeamrad)*j_mb_inframe_pm(2)

\begin{eqnarray}
\hat{j}_{bf} & = & [-{\rm sin}(\psi_B) \hat{i'}_{bf,x} + {\rm cos}(\psi_B) \hat{j'}_{bf,x}] \hat{i}
              + [-{\rm sin}(\psi_B) \hat{i'}_{bf,y} + {\rm cos}(\psi_B) \hat{j'}_{bf,y}] \hat{j} \nonumber \\
             & ~ &  + [-{\rm sin}(\psi_B) \hat{i'}_{bf,z} + {\rm cos}(\psi_B) \hat{j'}_{bf,z}] \hat{z} \, ,
\end{eqnarray}

%        jj_mb_inframe_pm(0)=-sin(psibeamrad)*i_mb_inframe_pm(0)
%     &                      +cos(psibeamrad)*j_mb_inframe_pm(0)
%        jj_mb_inframe_pm(1)=-sin(psibeamrad)*i_mb_inframe_pm(1)
%     &                      +cos(psibeamrad)*j_mb_inframe_pm(1)
%        jj_mb_inframe_pm(2)=-sin(psibeamrad)*i_mb_inframe_pm(2)
%     &                      +cos(psibeamrad)*j_mb_inframe_pm(2)

\noindent
where the bottom index $x$ $(y,z)$ indicates the component of 
{\it intermediate frame} unit vector along 
the axis $x$ $(y,z)$ of the {\it telescope frame}, as defined by 
eqs.~(A1--A3).

\end{document}